\documentclass{article}
\pdfoutput=1 % to allow compilation in JCAP
\usepackage{jcappub}

\usepackage{amsmath}
\usepackage{bm}
\usepackage{graphicx}
\usepackage{enumitem}
\usepackage{geometry}
\usepackage{xspace}
\usepackage{pdflscape}
\usepackage{placeins}
\usepackage{epstopdf}
\usepackage{multicol}
\usepackage{comment}
\usepackage{caption}
\usepackage{subcaption}
\usepackage{braket}
\usepackage{color}

\usepackage{multirow}
\usepackage{array}
\usepackage{longtable}

\makeatletter
\gdef\@fpheader{}
\g@addto@macro\bfseries{\boldmath}
\makeatother

\newcommand{\Mp}{M_\mathrm{Pl}}
\newcommand{\dd}{\mathrm{d}}
\newcommand{\zetan}{\zeta_n}
\newcommand{\Rpi}{\mathcal{R}_{\pi_0}}
\newcommand{\barP}{\bar{\mathcal{P}}}

 \def\be   {\begin{equation}}   \def\ee   {\end{equation}}
 \def\ba   {\begin{array}}      \def\ea   {\end{array}}
 \def\bea  {\begin{eqnarray}}   \def\eea  {\end{eqnarray}}
 \def\bean {\begin{eqnarray*}}  \def\eean {\end{eqnarray*}}

\definecolor{verde}{rgb}{0,0.5,0}
\definecolor{bordeaux}{rgb}{0.5, 0, 0.12}

\subheader{}

\title{Primordial Stochastic Gravitational Wave Background Anisotropies:\\ in-in Formalization and Applications}

\author[a,b]{Ema Dimastrogiovanni,}
\author[c,d]{Matteo Fasiello,}
\author[c]{Lucas Pinol}

\affiliation[a]{Van Swinderen Institute for Particle Physics and Gravity, University of Groningen, Nijenborgh 4, 9747 AG Groningen, The Netherlands}
\affiliation[b]{School of Physics, The University of New South Wales, Sydney NSW 2052, Australia}
\affiliation[c]{Instituto de F\'{i}sica T\'{e}orica UAM-CSIC, Calle Nicolás Cabrera 13-15, Cantoblanco, 28049,
Madrid, Spain}
\affiliation[d]{Institute of Cosmology \& Gravitation, University of Portsmouth, PO1 3FX, UK}

\emailAdd{lucas.pinol@ift.csic.es}

\date{today}

\begin{document}

\sloppy

\abstract{Primordial non-Gaussianities of the scalar(tensor)-tensor-tensor type supporting a non-trivial squeezed component are known to induce anisotropies in the stochastic gravitational wave background.
We derive the explicit form of such anisotropies by making use, for the first time in this context, of the in-in formalism for cosmological correlation functions.
After illustrating the general method and using it
for the minimal single-field slow-roll case, we apply it to multi-field models, providing both a tree-level and a one-loop example.
First, we make contact with previous results on anisotropies due to the presence of an extra spin-2 field during inflation.
Secondly, we calculate the 1-loop scalar-tensor-tensor three-point function in the context of so-called supersolid inflation.
The corresponding gravitational wave anisotropy is induced atop a gravitational signal that may be sufficiently large for detection.} 

\keywords{physics of the early universe, inflation, primordial stochastic gravitational wave background, laser interferometers}

% \arxivnumber{XXXX.XXXXX}

\maketitle
\section{Introduction}
\label{sec:intro}
The launch of future gravitational-wave (GW) detectors such as LISA \cite{LISA:2017pwj}, the Einstein Telescope \cite{Maggiore:2019uih}, and the planned BBO/DECIGO \cite{Kawamura:2006up,Corbin:2005ny}, will widen the frequency range over which we may test gravitational waves and, with them, the physics of the early universe. Their improved sensitivity with respect to existing probes of gravity will reveal a treasure trove of information on primordial mechanisms of GW generation. Combined with CMB polarization experiments such as CMB-S3\&4 \cite{CMB-S4:2020lpa} and LiteBIRD \cite{Matsumura:2013aja}, as well as pulsar timing arrays (EPTA, SKA), these will enable access to more than twenty decades in frequency, holding the potential to transform both cosmology and particle physics. 

Inflation is a case in point. Gravitational waves are a universal prediction of the inflationary dynamics but the specific features of the signal vary wildly across different realizations. The single-field slow-roll (SFSR) paradigm typically supports a slightly red-tilted GW spectrum, making detection at small scales much harder than that of B-modes of the CMB. Both the CMB-S4 and LiteBIRD experiment will be able to reach a sensitivity of $\sigma_r \sim 10^{-3}$ on the tensor-to-scalar ratio $r$ on large scales, with a guaranteed ruling in (out) of compelling models such as Starobinsky and Higgs inflation\footnote{The corresponding prediction on $r$ are the same in the large field limit \cite{Kehagias:2013mya}.}. 

The possibility of a GW detection at CMB scales is particularly intriguing in the case of the minimal SFSR scenario because in such set-up a one-to-one  relation connects the tensor to scalar ratio to the energy scale  of inflation\footnote{Strictly speaking, $E_{\rm inf}=\sqrt{H M_{\rm P}}$, but it is standard parlance to draw the correspondence directly with the Hubble rate.} $H$. In this case, a GW detection would turn each and every inflationary observable into a portal to a specific, and likely very high, energy scale: this is the ideal testing ground for beyond-Standard-Model physics and, possibly, quantum gravity.
On the other hand, extraordinary claims require extraordinary evidence, the latter taking the form of a thorough characterization of the primordial GW signal across all accessible scales.

A primordial GW detection at small scales can be equally informative: it would be strongly suggestive of a multi-field (or multi-clock) inflationary mechanism, one sufficiently different from the SFSR hypothesis to support, for example, a blue spectrum. More broadly, detection at intermediate (i.e. in the PTA range) and (or) small  scales (at or above LISA frequencies) may present with distinct features in the GW spectrum that enable a precise map between theory and data.

The need for a complete characterization of the GW signal, including frequency profile, chirality, and non-Gaussianity is then necessary to chart a clear map from observations to theory space. The study (and observation) of GW anisotropies at small and intermediate scales represents an additional, key, handle on the GW signal.  Indeed, direct access to primordial GW non-Gaussianities in such regime is hindered by propagation effects \cite{Bartolo:2018evs}: the initial correlation among different\footnote{For a direct measurement of the bispectrum, modes must stay within a precise k-interval, e.g. in the LISA band in the case of that specific instrument.} GW modes is washed out by their different history, i.e. by the different path GWs travel through structure to reach a given GW detector. Remarkably, there exists a specific configuration that does not suffer from such suppression of the signal: the squeezed ($k_3\ll k_2\sim k_1$) bispectrum limit. It is intuitively clear how a very long mode (up to horizon size) is much less sensitive to propagation through structure, whilst two nearly identical short modes share a very similar history. GW anisotropies are sensitive to precisely this bispectrum configuration \cite{Dimastrogiovanni:2019bfl}, sometimes called ultra-squeezed. Besides modulation from a long tensor mode, GW anisotropies emerge  also as the effect of a long scalar mode. It is on this last possibility that we shall mostly focus for the purposes of this work: testing inflationary interactions at small scales by their induced anisotropy on the GW power spectrum.

The result for the induced anisotropy was first written down in \cite{Jeong:2012df,Dai:2013kra,Brahma:2013rua,Dimastrogiovanni:2014ina,Dimastrogiovanni:2015pla} and later extended and used in \cite{Dimastrogiovanni:2021mfs},
both for the scalar-tensor-tensor (STT) \cite{Adshead:2020bji,Malhotra:2020ket} and the purely gravitational (TTT) cases \cite{Dimastrogiovanni:2019bfl}. In this work, we would like to provide a derivation of such results using the in-in formalism \cite{Schwinger:1960qe,Jordan:1986ug,Calzetta:1986ey,Weinberg:2005vy}. The latter offers more control over the derivation and all the underlying assumptions. Further, its starting point is directly the inflationary (quadratic and cubic) Lagrangian, from where one can proceed step by step to calculate the relevant cosmological correlation functions. We clarify both the assumptions implicit in previous literature on GW anisotropies and the precise sub-set of Feynman diagrams to which such results apply. To illustrate the power of the formalism we provide three examples. 

First, we focus on the minimal SFSR scenario. The corresponding signal is, as well-known, highly suppressed but such example will nevertheless clarify the essential aspects of the calculation whilst avoiding the complications seen in models with a richer dynamics. We then briefly consider the case of an inflationary theory equipped with an additional spin-2 field non-minimally coupled to the inflaton~\cite{Bordin:2018pca}, and show formally how the anisotropies are arrived at by using in-in techniques. Last, we consider in detail the case of supersolid inflation~\cite{Celoria:2020diz,Celoria:2021cxq,Comelli:2022ikb}. We compute for the first time the one-loop STT bispectrum and show how it features in the derivation of the corresponding GW anisotropies. 

The last two inflationary models are particularly interesting in that they support\footnote{In the case of the EFT with an extra spin-2, a detectable blue GW is contingent upon considering a time (or scale) dependent sound speed for the helicity-2 mode \cite{Iacconi:2020yxn}.} a detectable GW spectrum at small scales, which is a pre-condition for GW anisotropies to provide a realistic handle on inflationary interactions. Crucially, there exist also GW anisotropies of a different nature, including those of the astrophysical background \cite{Cusin:2017fwz,Cusin:2018rsq,Jenkins:2018kxc,Cusin:2019jhg,Bertacca:2019fnt,Pitrou:2019rjz,Bellomo:2021mer} and those due to propagation effects \cite{Contaldi:2016koz,Bartolo:2019oiq,Domcke:2020xmn,Alba:2015cms,Braglia:2021fxn}, and these ones too must be accounted for. We shall work under the assumption (realized in the 2 multi-field models under scrutiny here) of a large,  $f_{NL}\gg 1$, squeezed non-Gaussianity, which ensures that intrinsic (i.e. non-Gaussianity-induced) anisotropies give a much larger contribution than their ``propagation'' counterpart. As far as the possible degeneracy with anisotropies of the astrophysical background is concerned, that is beyond the scope of this work, but we stress that the
disentangling power of cross-correlations introduced in \cite{Adshead:2009cb,Ricciardone:2021kel,Dimastrogiovanni:2021mfs} applies precisely in cases such as the ones studied here.\\

This paper is organized as follows: in \textit{Section} \ref{section1} we introduce the in-in formalism in the context of the calculation of GW anisotropies and briefly apply it to the single-field slow-roll case; in \textit{Section} \ref{section_spin_2} we consider the case of spin-2 fields non-minimally coupled to the inflaton; \textit{Section} \ref{suso} is devoted to studying the STT bispectrum and corresponding anisotropy in the supersolid model; a discussion of the results and comments on future work are to be found in \textit{Section} \ref{conclusions}; \textit{Appendix} \ref{ininstandard} serves as a reminder for the usual in-in formalism; \textit{Appendix} \ref{app: loop calculations}    
provides more details on the calculations in supersolid inflation.

\section{Anisotropies: in-in formalisation}
\label{section1}
\subsection{Empiric formula for gravitational wave anisotropies}
The notion of  anisotropies induced by  primordial long-short mode coupling has first been introduced in \cite{Jeong:2012df,Dai:2013kra,Brahma:2013rua,Dimastrogiovanni:2014ina,Dimastrogiovanni:2015pla}, where anisotropies of the scalar power spectrum were considered. The extension of these results to the tensor two-point function was derived in \cite{Dimastrogiovanni:2019bfl} (see also \cite{Ricciardone:2017kre}). The proposed formula to take into account non-Gaussianity-induced (often referred to as ``intrinsic'') anisotropies is:

\bea
\label{q1}
&&\langle  \hat{\gamma}_{\textbf{k}_{1}}^{\lambda_{1}}\hat{\gamma}_{\textbf{k}_{2}}^{\lambda_{2}}\rangle \Big|_{\gamma_{L}}
\,
\equiv\, (2\pi)^{3}\,\delta^{\lambda_{1}\lambda_{2}}\,\delta^{(3)}(\textbf{k}_{1}+\textbf{k}_{2})P^{\lambda_{1}}_{\gamma}(k_{1})+\\
&& \sum_{\lambda_{3}}\int_{|\vec q|<q_L}d^{3}q\,\delta^{(3)}(\textbf{k}_{1}+\textbf{k}_{2}+\textbf{q}) \gamma^{*\lambda_{3}}_{\textbf{q}}\,\frac{B_{\gamma}^{\lambda_1 \lambda_2 \lambda_3}(\textbf{k}_{1},\textbf{k}_{2},\textbf{q})}{P^{\lambda_{3}}_{\gamma}(q)}\,,
\nonumber
\eea
where the case of a TTT three-point function has been considered. The generalization to the STT bispectrum is straightforward. The quantities in Eq.~(\ref{q1}) are the tensor power spectrum $P^{\lambda}_{\gamma}$ with $\lambda$ polarization index, and the TTT bispectrum is indicated by $B_{\gamma}^{\lambda_1 \lambda_2 \lambda_3}(\textbf{k}_{1},\textbf{k}_{2},\textbf{q})$. Note that the domain of integration $|\vec{q}|<q_L$ serves as a reminder that the squeezed  configuration, $q\ll k_1 \simeq k_2$, is the one under study and $\gamma_L$ stands for a long wavelength tensor mode.

We now set to derive this formula from first principles by using the in-in formalism. In the process we will clarify the assumptions underlying the result in Eq.~(\ref{q1}). We apply our results to specific inflationary models in Sections \ref{SFSR specific}, \ref{section_spin_2} and \ref{suso} of the text. We refer the reader to Appendix \ref{ininstandard} for a very brief introduction on the in-in formalism. In the main text we will consider directly its application to the two-point function in the presence of long-short mode coupling.

\subsection{Tensor two-point function in the presence of a long mode}
In this work, we shall be interested in computing the anisotropies of the stochastic gravitational wave background, at frequencies probed by next-generation GW experiments, due to the modulation from a very long wavelength mode.
One key aspect of our description of the long wavelength mode modulation on the power spectrum will be the possibility of treating the soft mode as classical.
This is standard practice in calculating scattering amplitudes \cite{Peskin:1995ev} and has seen plenty of applications in the inflationary context, for example in clarifying\footnote{Note that the parallel here is only on the fact that the long mode is treated classically. The last two models considered in this work actually break  consistency relations.} the physics of the so-called consistency relations \cite{Creminelli:2004yq,Ganc:2010ff,Renaux-Petel:2010paw,Hinterbichler:2013dpa}.

In the context of a quantum theory, as it is the case for primordial perturbations during inflation, it is possible to implement a classical treatment for a source on large scales.
Operationally, within a given interaction Hamiltonian $\hat{H}_\mathrm{int}=\int \dd^3 \vec{x} \, \hat{\mathcal{H}}_\mathrm{int}$, one of the quantum fields may well be approximated as a real-valued stochastic variable with negligible gradients.
Symbolically, if $\hat{\mathcal{H}}_\mathrm{int} \supset \hat{h}_\mathrm{int} \times \hat{J}$, where $\hat{h}_\mathrm{int}=\hat{h}_\mathrm{int}(\hat{\psi}_i)$ is a composite quantum operator made of fundamental fields, one can consider the limit in which $\hat{J}$ is treated as a classical source  $J^\mathrm{cl}$, under the condition that $J^\mathrm{cl}$ acts as a background, i.e. that $\partial_i J^\mathrm{cl} \ll \partial_i h_\mathrm{int}$. { 
Such a classical source is nevertheless an operator, being a classical random variable: physical observables are made of ensemble averages.
In particular, light particles during inflation are well approximated by massless quantum fluctuations $\hat{Q}_{\vec{k}}$ in de Sitter and, upon crossing the Hubble radius, undergo such classicalisation~\cite{PhysRevD.42.3413,Albrecht_1994,Polarski:1995jg,KIEFER_1998,SUDARSKY_2011,Martin_2016,Ashtekar:2020mdv} :
\begin{align}
\label{eq: classicalisation}
    \hat{Q}_{\vec{k}}(\tau) \,\, = \quad & Q_k(\tau) \hat{a}_{\vec{k}} +  Q_k^*(\tau) \hat{a}^\dagger_{-\vec{k}}\,, \quad \text{ with } \,\, Q_k(\tau) \propto (1+ik\tau)e^{-i k \tau} \nonumber \\
    \underset{-k \tau \rightarrow 0}{\longrightarrow} & \, Q^\mathrm{cl}_{\vec{k}}(\tau) = Q_k(\tau) \left(\hat{a}_{\vec{k}} +\hat{a}^\dagger_{-\vec{k}} \right) \,,
\end{align}
where conformal time $\tau$ is being used for convenience, $\dd t = a\,\dd \tau$.
The factorisation of the annihilation and creation operators is made possible by the fact that the mode function $Q_k(\tau)$ asymptotes to a real-valued quantity on super-horizon scales.
The operator $b_{\vec{k}}=\hat{a}_{\vec{k}} +\hat{a}^\dagger_{-\vec{k}}$ then is the only one appearing in the super-horizon limit for $\hat{Q}_k$ and therefore has no non-trivial commutation relation: it is a classical random variable.
This understanding of light fields, quantised at small scales but treated as classical on super-horizon scales, is also at the heart of the stochastic formalism for inflation (see, e.g.~\cite{Starobinsky:1986fx,NAMBU1988441,Starobinsky:1994bd} for pioneering works and~\cite{Pinol:2020cdp} for a more recent perspective).
Note also that for sufficiently light particles, the time dependence of the super-horizon variable is suppressed, in which case one can simply consider it as a constant random variable~\cite{Pinol:2020cdp}:
\begin{equation}
    Q^\mathrm{cl}_{\vec{k}}(\tau) \simeq Q^\mathrm{cl}_k \times b_{\vec{k}} \,, \quad \text{with} \,\, \braket{b_{\vec{k}}}=0 \,, \quad \braket{b_{\vec{k}}b_{\vec{k}^\prime}}=(2\pi)^3 \delta^{(3)}(\vec{k}+\vec{k}^\prime) \quad \text{and} \,\, Q^\mathrm{cl}_k = \underset{-k\tau \rightarrow 0}{\mathrm{lim}}  Q_k(\tau) \,. 
\end{equation}
In practice, this is the ansatz we shall use throughout this article to treat the classical source $J^\mathrm{cl}$.}

The simplest possibility in the context of gravitational wave anisotropies is that of a tree-level cubic interaction of the two tensor modes with the classical source $J^\mathrm{cl}$: 
\begin{align}
\label{eq210}
    \int \dd t \hat{H}_\mathrm{int} =  \int \dd \tau a(\tau) \int\frac{\dd^3 \vec{k} \dd^3 \vec{q}}{(2\pi)^6}D\Big[ \hat{\gamma}_{\vec{k}}(\tau) \hat{\gamma}_{\vec{q}}(\tau)\Big] J^\mathrm{cl}_{-\vec{k}-\vec{q}}(\tau)\,,
\end{align}
where $D\big[\cdots\big]$ is a placeholder for the polarization tensors as well as, crucially, time and/or space derivatives acting on the tensor modes\footnote{
It is also in general a function of the scale factor $a(\tau)$ as well as slowly-varying background quantities.}
 and will depend on the specific model at hand.
By replacing this interaction (see the right panel of Fig~\ref{fig: direct cubic} for the corresponding Feynman diagram) in Eq.~(\ref{formula1}), one obtains
\begin{align}
\label{eq211}
    \Braket{\hat{\gamma}^\lambda_{\vec{k}_1}\hat{\gamma}^{\lambda^\prime}_{\vec{k}_2}}^{J^\mathrm{cl}} =  - 2 \mathrm{Im}
    \left[\int_{-\infty^+}^0 \dd \tau^\prime a(\tau^\prime)    \int\frac{\dd^3 \vec{k} \dd^3 \vec{q}}{(2\pi)^6}  J^\mathrm{cl}_{-\vec{k}-\vec{q}}(\tau^\prime) \Braket{0| D\Big[\hat{\gamma}_{\vec{k}}(\tau^\prime) \hat{\gamma}_{\vec{q}}(\tau^\prime)\Big]  \hat{\gamma}^\lambda_{\vec{k}_1}\hat{\gamma}^{\lambda^\prime}_{\vec{k}_2} |0} \right] \,,
\end{align}
where the superscript $^I$, has been removed  for simplicity.
The classical source $J^\mathrm{cl}$ has been factored out of the vacuum expectation value since it is not a quantum operator. The next step is to show that, under mild assumptions, the classical source can be also taken out of the time integral.

Once quantum operators are contracted, the source $J$ has momentum $-\vec{k}_1-\vec{k}_2$. In the quasi-diagonal configuration one has $|\vec{k}_1+\vec{k}_2|\equiv k_L\ll k_1,k_2\simeq k_S$. Then, if $J$ is well-approximated by a constant for small values of the argument (i.e. for sufficiently late times $-k_L\tau\ll1$), and the useful time domain of integration in Eq.(\ref{eq211}) only extends to such time interval,  $J^{\rm cl}$ can indeed be factored out completely.
This is typically the case of the integral in Eq.(\ref{eq211}) whenever the squeezed limit hierarchy $k_L \ll k_S$ is applied.
Indeed, the tensor mode functions feature a $\gamma(k_S\tau)$ dependence
\footnote{Note that this line of reasoning extends also to the case when  such functions do not have the typical $H_{3/2}^{(1)}(k_S\tau)$ argument where $H_\nu^{(1)}$ stands for an Hankel function of the first type.
It might be the case that, through direct coupling at the level of the quadratic Lagrangian with other tensor degrees of freedom, the $\gamma$ wavefunctions inherit a massive component (i.e. a component proportional to $H_{\nu}^{(1)}(k_s\tau)$ with $\nu< 3/2$).
It is also possible that $\gamma$ functions exhibit a sound speed $c_{s}$ dependence which would move the horizon of the wavefunction from $-k_S \tau\sim 1$ to $-c_s k_S \tau\sim 1$.
In all these cases it is still true that the (by far) leading contribution to cosmological correlation functions, and in particular to the integral in Eq.~(\ref{eq211}) comes nevertheless from integrating over the domain satisfying $-(c_s)\,k_S\,\tau\lesssim 1 $.
For our reasoning to be valid in these more general scenarios, it is sufficient to ask that $k_L \ll c_s\,k_S$, which is always the case in the ultra-squeezed configuration we shall pursue.},
which immediately restricts the useful time domain of integration to the $-k_S\tau\lesssim 1$ region. This is due to the fact that the $\gamma$ modes are fast oscillating in the complementary region, that is, inside their horizon.
The hierarchy then enforces the condition $-k_L\tau\ll - k_S\tau\lesssim 1$. One can therefore safely consider the zeroth-order approximation:

\begin{equation}
    J^\mathrm{cl}_{\vec{k}_1+\vec{k}_2}(\tau) \simeq  J^\mathrm{cl}_{\vec{k}_1+\vec{k}_2}(0)\,.
\end{equation}
The tensor two-point function in the presence of a classical, real, source $J^\mathrm{cl}$ then takes the following form:
\begin{equation}
\label{eq: power spectrum with classical source}
    \Braket{\hat{\gamma}^\lambda_{\vec{k}_1}\hat{\gamma}^{\lambda^\prime}_{\vec{k}_2}}^{J^\mathrm{cl}} \underset{|\vec{k}_1+\vec{k}_2| \ll k_1, k_2}{=}
    - 4  J^\mathrm{cl}_{|\vec{k}_1+\vec{k}_2|} \mathrm{Im}
    \left\{ \gamma^{\lambda*}_{k_1}\hat{\gamma}^{\lambda^{\prime} *}_{k_2}
    \int_{-\infty^+}^0 \dd \tau^\prime a(\tau^\prime) 
    D\Big[ \gamma^\lambda_{k_1}(\tau^\prime) \gamma^{\lambda^{\prime}}_{k_2}(\tau^\prime)\Big]   \right\}\,,
\end{equation}
which shows how different values of $J^\mathrm{cl}$ on large scales $|\vec{k}_1+\vec{k}_2| \ll k_1, k_2$  affect the tensor two-point function on smaller scales, thereby inducing  anisotropies in the SGWB.

\subsection{Squeezed bispectrum and GW anisotropies}
\begin{figure}
    \centering
    \begin{subfigure}{0.48\textwidth}
        \centering
        \includegraphics[width=1.\linewidth]{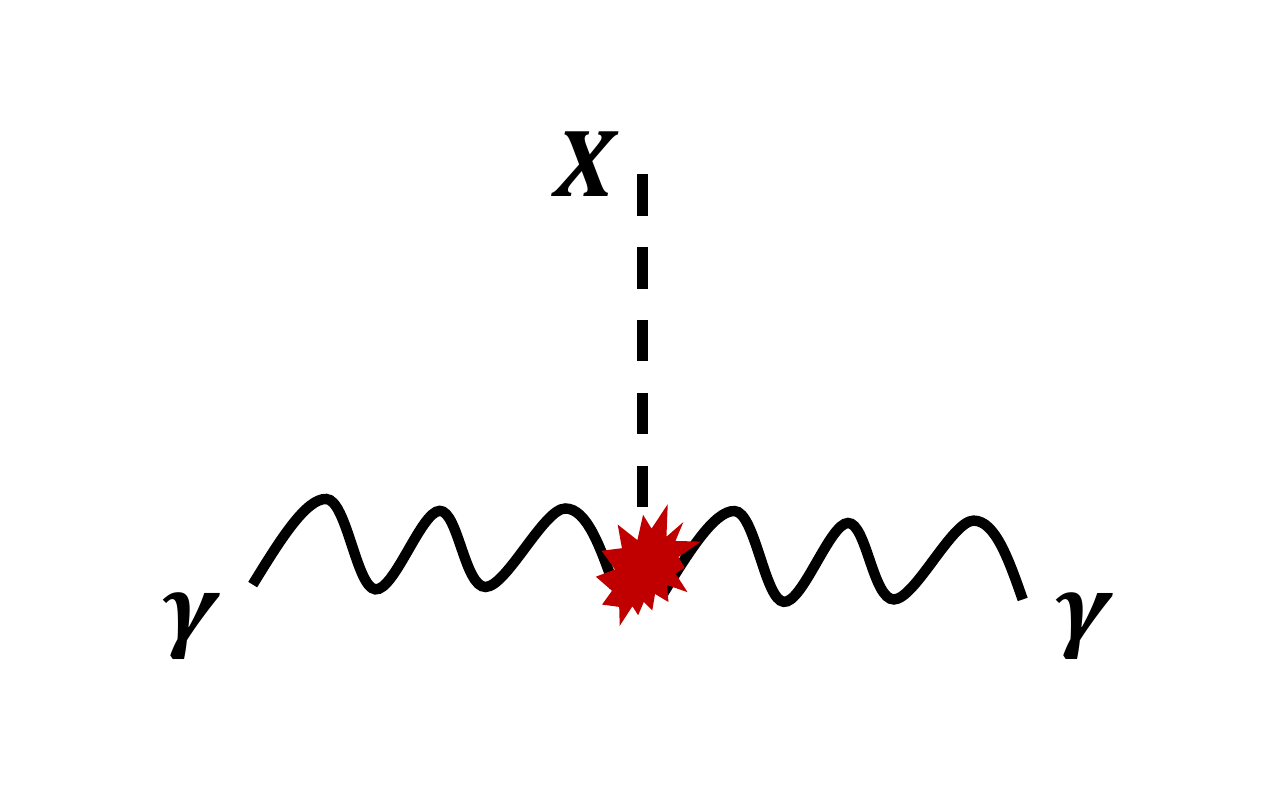}
        \caption{Mixed X-tensor-tensor bispectrum.}
    \end{subfigure}%
    \hfill
    \begin{subfigure}{0.48\textwidth}
        \centering
        \includegraphics[width=1.\linewidth]{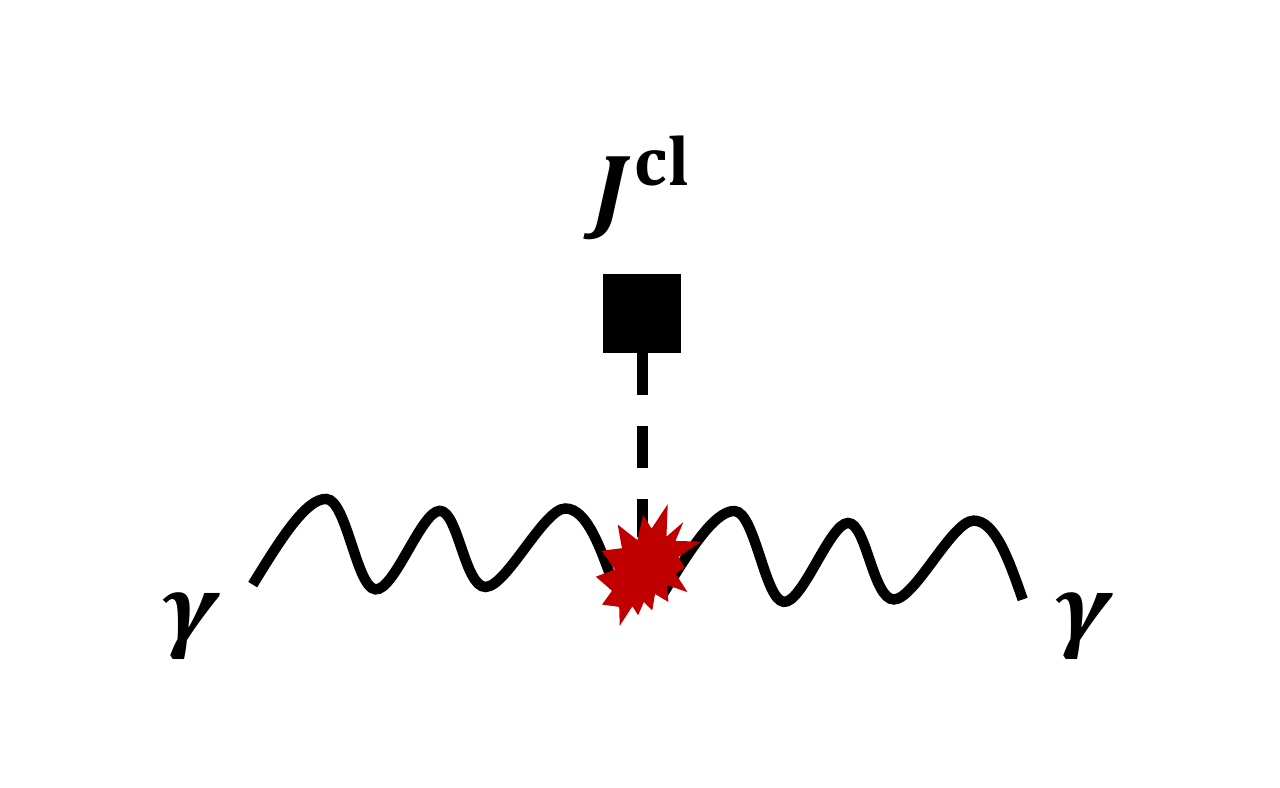}
        \caption{Tensor two-point function in the presence of a classical scalar source $J^\mathrm{cl}$.}
    \end{subfigure}
    \caption{Relevant Feynman diagrams for the primordial anisotropies in the case of a direct cubic coupling. A straight dashed line represents a mixed $J$-$X$ propagator, a wavy line represents a $\gamma$ propagator. The square at the top of the dashed line represents, in contradistinction to a propagator connecting to an external leg, a classical background $J^\mathrm{cl}$.}
    \label{fig: direct cubic}
\end{figure}
We would like to arrive at Eq.~\eqref{q1} by means of the in-in formalism, making explicit contact with the bispectrum of a given theory. This shall be  possible in light of the particular kinematic configuration of the squeezed limit.
Let us look at the three-point function between two tensor modes $\hat{\gamma}$, and another quantum field $\hat{X}$ that can be contracted together with the quantum version of the classical source, $\hat{J}$, i.e. with a non-trivial commutation relation between the two:
\begin{equation}
    \left[\hat{J}_{\vec{k}}(\tau),\hat{X}_{\vec{k}^{\prime}}(\tau^\prime)\right] = (2\pi)^3 \delta^{(3)}(\vec{k}+\vec{k}^\prime) \left[J_k(\tau) X_{k^{\prime}}^*(\tau^\prime) - J_k^*(\tau) X_{k^{\prime}}(\tau^\prime)\right].
\end{equation}
Note that in many examples $J$ and $X$ are the very same field. However, we are considering here a more general case in order to set the stage for the applications in the coming sections. 

Focusing on the same cubic interaction as in Eq.(\ref{eq210}) leads, in the case of the bispectrum, to the computation of the Feynman diagram in the left panel of Fig.~\ref{fig: direct cubic}, where:
\begin{align}
\label{}
    \int \dd t \hat{H}_\mathrm{int} = \int \dd \tau a(\tau)  \int\frac{\dd^3 \vec{k} \dd^3 \vec{q}}{(2\pi)^6}D\Big[ \hat{\gamma}_{\vec{k}}(\tau) \hat{\gamma}_{\vec{q}}(\tau)\Big] \hat{J}^\mathrm{}_{-\vec{k}-\vec{q}}(\tau)\,.
\end{align}
We can express the corresponding three-point function as:
\begin{align}
    \Braket{\hat{\gamma}^\lambda_{\vec{k}_1}\hat{\gamma}^{\lambda^\prime}_{\vec{k}_2} \hat{X}_{\vec{k}_3}}
    =  - 2 \mathrm{Im}&\left[\int_{-\infty^+}^0 \dd \tau^\prime a(\tau^\prime)   \int\frac{\dd^3 \vec{k} \dd^3 \vec{q}}{(2\pi)^6}  \right.  \\
    & \times \Braket{0| D\Big[\hat{\gamma}_{\vec{k}}(\tau^\prime) \hat{\gamma}_{\vec{q}}(\tau^\prime) \Big]
    \hat{J}_{-\vec{k}-\vec{q}}(\tau^\prime) \hat{\gamma}^\lambda_{\vec{k}_1}
    \hat{\gamma}^{\lambda^\prime}_{\vec{k}_2}
    \hat{X}_{\vec{k}_3}
    |0} \Bigg] \,, \nonumber
\end{align}
where we are going to consider the contraction of  $\hat{J}$ with $\hat{X}$ (i.e. the left diagram of Fig.~\ref{fig: direct cubic}). In the limit where the external mode $X$ is very soft, i.e. in the configuration $k_3 \ll k_1, k_2$, one may implement the same simplifying steps  we used above.
Indeed, after performing the Wick contractions, the argument of the mode function $J_{k_3}(\tau^\prime)$ is already very small at the time of horizon crossing for the tensor modes with wavenumbers $k_1, k_2$.
Recall that it is the latter that set the effective lower extremum of integration over the time domain.
It follows that both $J$ and $X$ can be factored out, yielding:
\begin{align}
\label{eq: mixed bispectrum squeezed}
    \sum_{\lambda,\lambda^\prime}\Braket{\hat{\gamma}^\lambda_{\vec{k}_1}\hat{\gamma}^{\lambda^\prime}_{\vec{k}_2}\hat{X}_{\vec{k}_3}} =
    & (2\pi)^3
    \delta^{(3)}(\vec{k}_1+\vec{k}_2+\vec{k}_3) 
    B^{\gamma \gamma X}(k_1,k_2,k_3) \,, \quad     \mathrm{with} \\
    B^{\gamma \gamma X}(k_1,k_2,k_3)
    \underset{k_3 \ll k_{1,2}}{=}  - 4 & P_{JX}(k_3) \times \mathrm{Im}\left\{
     \sum_{\lambda,\lambda^\prime}
     \gamma^{\lambda*}_{k_1}\hat{\gamma}^{\lambda^{\prime} *}_{k_2}
     \int_{-\infty^+}^\tau \dd \tau^\prime a(\tau^{\prime})
    D\Big[ \gamma^\lambda_{k_1}(\tau^\prime) \gamma^{\lambda^{\prime}}_{k_2}(\tau^\prime) \Big]  \right\}\,, \nonumber
\end{align}
where we defined the cross power spectrum $P_{JX}$ as
\begin{equation}
\label{eq218}
    \underset{\tau \rightarrow 0}{\mathrm{lim} }\Braket{0|\hat{J}_{\vec{k}}(\tau)\hat{X}_{\vec{k}^\prime}(\tau)|0} = (2\pi)^3 \delta^{(3)}(\vec{k}+\vec{k}^\prime) P_{JX}(k) \,.
\end{equation}
We stress that $P_{JX}$ may well be a standard power spectrum if $X=J$. It is also possible, as we shall see, that the two fields are coupled at the level of the quadratic Lagrangian. In the limit where  such coupling cannot be treated as a perturbation $\delta\mathcal{L}_2$ of a free Lagrangian $\mathcal{L}_2$, one may still solve the system non-perturbatively (see e.g. the Appendix of \cite{Bordin:2018pca} for an interesting example). In such case the two fields will have a non-zero cross-correlation already at the linear level, leading to the general formula in  Eq.~(\ref{eq218}).

The resemblance of Eq.~\eqref{eq: mixed bispectrum squeezed} with Eq.~\eqref{eq: power spectrum with classical source} for the anisotropic tensor two-point function in the presence of a classical source, is apparent.
It is therefore straightforward to derive the formula:
\begin{equation}
\label{equa221}
     \sum_{\lambda,\lambda^\prime}\Braket{\hat{\gamma}^\lambda_{\vec{k}_1}\hat{\gamma}^{\lambda^\prime}_{\vec{k}_2}}^{J^\mathrm{cl}} \underset{|\vec{k}_1+\vec{k}_2| \ll k_1, k_2}{=} 
     \frac{B^{\gamma \gamma X}(k_1,k_2,|\vec{k}_1+\vec{k}_2|)}{P_{JX}(|\vec{k}_1+\vec{k}_2|)} J^\mathrm{cl}_{|\vec{k}_1+\vec{k}_2|} \,,
\end{equation}
which is valid for a generic field $J$ so long as it is correlated with the external field $X$ and one is probing the kinematic configurations $|\vec{k}_1+\vec{k}_2| \ll k_1, k_2$.
This expression, derived within the in-in formalism under a number of conditions, matches those already present in the literature but whose derivation is heuristic, with  $J=X$ being either a scalar mode $\zeta$ \cite{Adshead:2009cb} or a tensor fluctuation $\gamma$ \cite{Dimastrogiovanni:2019bfl}.

Although not the main focus of this work, which is on the in-in derivation of the effect of long-short coupling on primordial correlators, we find it useful to briefly report here on how this effect impacts gravitational wave anisotropies $\delta_{\rm GW}$. We shall use the results published in \cite{Dimastrogiovanni:2021mfs}, and refer the interested reader to \cite{Dimastrogiovanni:2021mfs} and references therein for a detailed treatment.
Starting from the energy density $\Omega_{\rm GW}(f)$, the definition of $\delta_{\rm GW}$ is:

\bea
\Omega_{\rm GW}(f)=\bar{\Omega}_{\rm GW}(f)\Big[1+\frac{1}{4\pi}\int \dd^2 \hat{n}\, \delta_{\rm GW}(f,\hat{n})  \Big]\; ,
\eea

\noindent where $f= k/2\pi$. In relatively simple\footnote{The generalization to the cases under scrutiny here is straightforward.} inflationary models (i.e. those where, in the language of Eq.~\eqref{equa221}, $X=J=\zeta$), one finds that:

\bea
\label{deltagw}
\delta^{\rm GW}_{STT}(k,\hat{n})=\int_{q \ll k} \frac{\dd^3 \vec{q}}{(2\pi)^3} e^{-i d\, \hat{n}\cdot \vec{q}}\,  f^{STT}_{\rm NL}\left(\vec{k},\vec{q}\right)\zeta\left(\vec{q}\right)\; ,
\eea
where $\vec{k}=k\, \hat{n}$. Note the condition on the domain of integration, ensuring compliance with the assumptions needed  (see e.g. Ref.~\cite{Dimastrogiovanni:2021mfs}) to arrive at Eq.~(\ref{deltagw}) and that the bispectrum configuration probed is the squeezed one.  The quantity $d= \eta_0 - \eta_{\rm in}$ is the conformal time elapsed from horizon-entry of the mode $k$ to the present. The quantity $f^{STT}_{\rm NL}$ is written in terms of the primordial bispectrum as:
\bea
f^{STT}_{\rm NL}\left(\vec{k},\vec{q}\right)= \frac{B\left(\vec{k}-\vec{q}/2, -\vec{k}-\vec{q}/2,\vec{q}\right)}{P_{\zeta}(q) P_{\gamma}(k)}\; .
\eea
A completely analogous relation is found for the GW anisotropy corresponding to a modulation due to a long tensor mode.

Gravitational wave anisotropies can serve as a useful handle on primordial non-Gaussianities at small scales. It is important to stress here that  this is an effective tool approximately up to multipoles of order 15-30, see e.g. \cite{Dimastrogiovanni:2021mfs,Malhotra:2020ket}. Indeed, previous studies \cite{Kudoh:2004he,Alonso:2020rar,Contaldi:2020rht} have shown that, for interferometers such as LISA, there are no improvements on $\sigma_{f_{\rm NL}}$ past those scales because the noise becomes important. One should also stress that, for the inflationary models of interest here, the modulation is more important for $\ell=2$ (see \cite{Dimastrogiovanni:2021mfs}).

\subsection{Single-field, slow-roll inflation.}
\label{SFSR specific}
We will now apply our results to the case of a TTT bispectrum and the GW anisotropy induced by such three-point function. This first and simplest example is that of single-field slow-roll (SFSR) inflation. We will be rather brief for two reasons. First, the result follows immediately from the procedure illustrated above. Secondly, the GW primordial power spectrum in SFSR is slightly red-tilted and thus unlikely (with the possible exception of BBO/DECIGO) to be detected  at intermediate and small scales by upcoming GW detectors. The TTT non-Gaussian anisotropy signal itself is very small, making detection of associated anisotropies unrealistic \cite{Dimastrogiovanni:2021mfs}. Furthermore, the fact that so-called consistency relations are in place in SFSR raises the issue of the leading contribution to the squeezed bispectrum being a gauge artifact. We nevertheless mention the SFSR GW bispectrum and anisotropy for the simplicity of the formal derivation and then move on to the more involved (but rather more interesting from the point of view of observations) multi-field scenarios. The tensor-tensor-tensor bispectrum in SFSR was first computed in \cite{Maldacena:2002vr,Maldacena:2011nz} (see also e.g. \cite{Gao:2011vs}). The interaction Hamiltonian in this case reads 
\begin{equation}
    \label{}
    \int \dd t {H}_\mathrm{int} =  \frac{\Mp^{2}}{2}\int \dd^{3} \vec{x} \int \dd \tau  \,
    a^2(\tau)\,\partial_{k}\partial_{l}\gamma_{ij}\left(\gamma_{ik}\gamma_{jl}-\frac{1}{2}\gamma_{ij}\gamma_{kl}\right) \,,
\end{equation}
and the corresponding three-point function in the squeezed limit is given by \cite{Maldacena:2002vr}
\begin{equation}
    \Braket{\hat{\gamma}^\lambda_{\vec{k}_1}\hat{\gamma}^{\lambda^\prime}_{\vec{k}_2}\hat{\gamma}^{\lambda_3}_{\vec{k}_3}}
    \underset{k_3 \ll k_1,k_2}{=}
    (2\pi)^3 \delta^{(3)}(\vec{k}_1+\vec{k}_2+\vec{k}_3)
    \delta^{\lambda_{}\lambda^{'}}\left(\frac{H}{\Mp}\right)^{4}\frac{3}{2}\frac{1}{k_{1}^{3}k_3^{3}}\epsilon_{ij}^{\lambda_{3}}(\hat{k}_3)\hat{k}_{1}^{i}\hat{k}_{1}^{j}\,.
\end{equation}
The results derived in this section then readily apply upon identifying $J_{\vec{k}_3}=X_{\vec{k}_3}=\gamma_{\vec{k}_3}^{\lambda_{3}}$.

\subsection{Multi-field scenarios}
\label{sec:models}

The use of anisotropies of the tensor two-point function as a probe of early universe physics applies to all scales, from CMB to high-frequency GW interferometers.
As we have seen in the case of single-field slow-roll (SFSR) inflation, a squeezed component of the primordial bispectrum induces an anisotropic component to the tensor two-point function.
At large scales (e.g. the CMB), one may directly access  non-Gaussianities, whilst GW anisotropies serve as an ancillary probe of the same physics. Crucially, this is no longer the case at intermediate and small scales\footnote{By ``intermediate'' here we shall mean scales probed by pulsar timing arrays (e.g. $f \simeq 10^{-9}$Hz), whilst we shall term ``small scales'' those accessible by laser interferometers, i.e. frequencies $f$ such that $f\gtrsim 10^{-3}$Hz.}. 

Indeed, let us consider e.g. initially correlated tensor modes that re-enter the horizon sufficiently late (e.g. during radiation domination) to be in the frequency range of laser interferometers. Inevitably\footnote{This follows directly from the fact that all $k$ modes need to be within a certain frequency band and from overall momentum conservation.}, all modes have to travel through structure to reach the detector and, in doing so, all undergo a different propagation history that effectively washes away any initial (i.e. primordial) correlation \cite{Bartolo:2018evs}. This same line of reasoning suggests one possible exception (see \cite{Powell:2019kid} for another interesting configuration): those modes where a very large hierarchy of scales is present. If it is the case, for example in a  three-point function, that $k_3\ll k_1\sim k_2$, then the two short modes will share a very similar history and the long one may undergo little propagation if e.g. horizon-size. Naturally, the non-Gaussianity corresponding to such momentum configuration, the ultra-squeezed one, may well not be tested directly, but it is precisely this configuration that we access when studying GW anisotropies. 

In order to make full use of such a handle on inflationary interactions, it is necessary that (i) the GW signal is accessible at small scales and (ii) the long-short mode coupling that induces the anisotropy is sufficiently large. Such requirements lead almost universally to a multi-field inflationary scenario, far away from the SFSR paradigm with its slightly red-tilted tensor power spectrum which is, at best, only accessible by the planned DECIGO/BBO experiment.

We shall see how to derive the STT bispectrum and associated GW anisotropies in two multi-field models of inflation: an EFT of non-minimally coupled extra spin-2 field \cite{Bordin:2018pca,Dimastrogiovanni:2018gkl,Iacconi:2019vgc,Iacconi:2020yxn} and supersolid inflation \cite{Ricciardone:2016lym,Celoria:2020diz}. The requirement (ii) of a non-trivial squeezed component rules out other classes of multi-field models with interesting GW phenomenology, such as those of axion (gauge fields) inflation (see  \cite{Pajer:2013fsa,Adshead:2012kp,Dimastrogiovanni:2016fuu,Domcke:2018rvv} and references therein), whose bispectrum shape is of the equilateral type \cite{Agrawal:2017awz,Agrawal:2018mrg,Dimastrogiovanni:2018xnn,Fujita:2018vmv}. Before detailing on the two specific inflationary set-ups, it is useful to recall a useful  ``rule of thumb'' for an educated guess on the so-called shape of the non-Gaussianity of a given model interaction: in the case of a Bunch-Davies vacuum, a squeezed bispectrum component is typically associated to non-derivative interactions of very light fields. Both of the examples we will focus on fall into this categorization.  

The most intuitive way to understand the line of reasoning behind the ``rule of thumb'' is to remember that a massive wave-function  typically comes with an extra factor of $(k \tau)$ w.r.t. its massless counterpart. This leads to a suppression in the $k\rightarrow 0$ region of $k$ space, to the detriment of the squeezed component. Similarly, time and space derivatives come with extra positive power of the momenta (as well as powers of $\tau$ in the case of time derivatives) with respect to non-derivative interactions, leading to the same effect. That being said, there is yet another aspect that is worth mentioning when it comes to the squeezed limit of primordial bispectra: consistency relations. 

Consistency relations stem for residual gauge diffeomorphisms in the description of a physical system. These relate, when applicable, e.g. the soft (that is, the squeezed) limit of a three-point function with the action of the residual gauge transformation on the corresponding (hard) two-point correlator. As such, the leading contribution of the squeezed bispectrum in a certain theory may be a gauge artifact. There is a rich literature~\cite{Maldacena:2002vr,Creminelli:2004yq,Hinterbichler:2013dpa,Creminelli:2013mca,Peloso:2013zw,Kehagias:2013yd} on consistency relations, a fascinating topic which is, to this day, the subject of an intense research activity~\cite{Matarrese:2020why}. For the purposes of this work, it will suffice to say that the interactions we will probe for both models under study do break consistency relations and lead therefore to unequivocally physical effects already at leading order. In the spin-2 case, the proof of consistency relation breaking is straightforward\footnote{It is a parametric proof, based on showing how certain cubic interactions are regulated by parameters that do not appear in the quadratic Lagrangian.} \cite{Bordin:2018pca,Iacconi:2019vgc} and completely analogous to the case of quasi-single field inflation \cite{Chen:2009we,Chen:2009zp,Assassi:2012zq}. Supersolid inflation, much like solid inflation, comes with a different symmetry breaking pattern w.r.t. standard scenarios (i.e. those where time-reparametrization is the only one broken by the background). This too leads to consistency relation breaking \cite{Endlich:2013jia,Bordin:2017ozj,Celoria:2020diz}.

By using the in-in formalism we shall derive the GW anisotropies in the case of an extra spin-2 field during inflation and make contact with the heuristic result already present in the literature \cite{Dimastrogiovanni:2021mfs}. Illustrating the power of anisotropies for supersolid inflation will require calculating both the one-loop STT bispectrum and the related GW anisotropy. We shall focus on the former in \textit{Section} \ref{section_spin_2} and detail on our results for supersolid inflation in \textit{Section} \ref{suso}.

\section{Non-minimally coupled spin-2 field}
\label{section_spin_2}
In exploring the rich particle content allowed by the multi-field inflationary paradigm, one may well consider extra\footnote{``Extra'' is meant in this context as in addition to the field content of the minimal scenario, that of general relativity as the theory of gravity plus a single scalar degree of freedom driving the acceleration.} scalar and vector fields \cite{Martin:2013tda}. Going further up the spin ladder, the next step is to posit the presence of spin-2 fields. Remarkably, even higher-spin fields have been considered in the literature \cite{Kehagias:2017cym,Baumann:2017jvh,Anninos:2019nib} and their signatures investigated \cite{Bartolo:2017sbu,MoradinezhadDizgah:2018pfo}.

Although spinning fields are those that give the most distinct signatures in terms, for example, of squeezed non-Gaussianities \cite{Kehagias:2015jha,Arkani-Hamed:2015bza}, unitarity constraints limit the allowed mass range for spin-2 (and higher) particles. Such bounds have the schematic form (and order of magnitude) $m\gtrsim H$. Now, a similarly massive field tends to decay within a few e-folds, to the detriment of any observational signature associated to its presence in the early universe. The presence of unitarity bounds stems from our understanding of particles as unitary irreducible representations of the spacetime isometry group. The isometries of de Sitter space dictate stringent constraints on the mass range of spin-2 particles. Similar results hold also for FLRW spacetime \cite{Fasiello:2012rw,Fasiello:2013woa}. Crucially, de Sitter isometries are broken by the inflationary background. It follows that coupling spinning particles directly to the (isometry-breaking) inflaton field will weaken the so-called Higuchi bound and allow for long-lived spin-2 particles.

This is part of the motivation behind the extension of the EFT approach to multi-field inflation to include a non-minimal coupling to the constant inflaton foliation \cite{Bordin:2017ozj}. We shall employ this set-up in this section and focus in particular on the presence of an extra (now allowed to be) light spin-2 field $\sigma$ coupled to the inflaton.
Note that the EFT approach of choice here is the one of an effective field theory of fluctuations around an FLRW background. The study of a full theory of interacting spin-2 fields during inflation is rather different, and it has been the subject of several studies (see e.g. \cite{Dimastrogiovanni:2018uqy,Goon:2018fyu,Dimastrogiovanni:2021cif}).

In this section we will use the in-in formalism to show the relation between GW anisotropies and STT bispectrum in the spin-2 case. The calculation of the induced anisotropy has been performed for this model only in the so called perturbative regime and only using the heuristic formula in Eq.~\eqref{q1}. Our results extend those in the literature in two directions: the derivation is via the rigorous in-in formalism and we are also able to tackle, at least formally, the non-perturbative regime.

In spelling out the various steps we shall point to and rely on a number of results in the literature and, whenever these are not essential for our purposes, we will refer the reader to such works for more details on certain specific findings. 

Our calculations will be done in two regimes. Firstly, we shall consider the case of fields coupled already at the linear level, with no extra assumption on the strength of the coupling. Although we will not  explicitly solve the equations of motion in this regime, we will nevertheless be able to formally arrive at the desired result in Eq.~(\ref{eq316}). As a second step, we will consider the regime where the coupling is small and can be treated perturbatively (i.e. as a two-vertex interaction, exactly as in quasi-single-field inflation).

\subsection{The model}

We consider the following Lagrangian \cite{Bordin:2017ozj} for inflationary fluctuations:
\begin{equation}
    \mathcal{L}=\mathcal{L}_\pi + \mathcal{L}_\gamma + \mathcal{L}_\mathrm{\sigma} + \mathcal{L}_\mathrm{mix} \,.
\end{equation}
$\mathcal{L}_\pi$ is the usual single-field slow-roll Lagrangian of the EFT of inflation~\cite{Cheung:2007st}, while $\mathcal{L}_\gamma$ describes the tensor modes of the spacetime metric in GR. With $\mathcal{L}_\sigma$ we indicate $\mathcal{L}_\sigma=\mathcal{L}_\sigma^{(2)}+\mathcal{L}_\sigma^{(3)}$, where $\mathcal{L}_\sigma^{(2)}$ is the free Lagrangian for the spin-2  field  and $\mathcal{L}_\sigma^{(3)}$ contains one key interaction term:
\begin{align}
\label{eq: Lsigma}
    \mathcal{L}_\sigma^{(2)} &= \frac{a^3}{4} \left[ \left( \dot{\sigma}_{ij}\right)^2 - \frac{c_2^2}{a^2}  \left(\partial_i \sigma^{jk} \right)^2  - \frac{3}{2a^2}(c_0^2 - c_2^2)  \left( \partial_i \sigma^{ij}\right)^2 - m^2 \left(\sigma^{ij}\right)^2 \right] \\
    \mathcal{L}_\sigma^{(3)} &= - a^3 \mu  \left( \sigma^{ij}\right)^3\,, \nonumber
\end{align}
with $\sigma^{ij}=\sigma_{(0)}^{ij}+\sigma_{(2)}^{ij}$ (we shall not consider vector modes), whose Fourier transforms can be decomposed into polarisation tensors:
\begin{align}
    \sigma_{(0)}^{ij}(\vec{k}) &=  \epsilon^{ij}_{(0)}(\hat{k}) \sigma_{(0),\vec{k}} \quad \text{and } \quad  \sigma_{(2)}^{ij}(\vec{k}) = \sum_{\lambda=+,\times} \epsilon^{ij}_\lambda(\hat{k}) \sigma_{(2),\vec{k}}^\lambda \,, \quad \text{with } \\
    \epsilon^{ij}_{(0)}(\hat{k})&=\sqrt{\frac{3}{2}}\left( \hat{k}^i \hat{k}^j - \frac{1}{3}\delta^{ij}\right) \,, \nonumber
\end{align}
$\epsilon^{ij}_\lambda(\hat{k})$ being the usual tensor polarisations.
Moreover, $\mathcal{L}_\mathrm{mix}=\mathcal{L}_\mathrm{mix}^{(2)}+ \mathcal{L}_\mathrm{mix}^{(3)}$ describes the mixed interactions, up to cubic order again, between the canonically normalised scalar and tensor modes of the spacetime metric, $\pi_\mathrm{c} = \sqrt{2\epsilon} H \Mp \pi $ and $\gamma_\mathrm{c} = \Mp \gamma$, and the  spin-2 field:
\begin{align}
    \mathcal{L}_\mathrm{mix}^{(2)} &= a^3 \left[ - \frac{\rho}{a^2 \sqrt{2\epsilon} H} \partial_i \partial_j \pi_\mathrm{c} \sigma^{ij} + \frac{1}{2} \rho \dot{\gamma}_{\mathrm{c},ij} \sigma^{ij} \right] \,, \\
    \mathcal{L}_\mathrm{mix}^{(3)} &= a^3 \left[ - \frac{\rho}{a^2 2\epsilon H^2 \Mp } \left(\partial_i \pi_\mathrm{c} \partial_j \pi_\mathrm{c} \dot{\sigma}^{ij} + 2 H \partial_i \pi_\mathrm{c} \partial_j \pi_\mathrm{c} \sigma^{ij}\right) + \frac{\tilde{\rho}}{a^2\epsilon H^2 \Mp} \dot{\pi}_{\mathrm{c}} \partial_i \partial_j \pi_\mathrm{c} \sigma^{ij} \right] \nonumber \,.
\end{align}
One can check that the quadratic Lagrangian above only mixes $\pi$ and $\sigma_{(0)}$ on one side, and $\gamma$ and $\sigma_{(2)}$ on the other. For ease of comparison with works employing a different notation, we also stress that the pseudo-Goldstone boson $\pi$ is related to the primordial curvature fluctuation via $\zeta = - H \pi$ \cite{Cheung:2007st}. 

\subsection{Quadratic Lagrangian}

Due to $\mathcal{L}_\mathrm{mix}^{(2)}$, $\zeta$ and $\gamma$ mix with the helicity modes of the spin-2 field already at the linear level. For an appropriate choice of the parameters such mixing is small and can be treated as a perturbation on top of the free Lagrangian: we investigate this possibility in Sec.~\ref{sec: perturbative spin2}. We shall first consider, formally,  the general case with coupled equations of motion such that one may not proceed to quantise the different fluctuations independently. One ought to consider the general decomposition:
\begin{align}
    \hat{\zeta}_{\vec{k}} &= \sum_{A=1}^2 \zeta_k^{A} \hat{a}_{\vec{k}}^{A} + \mathrm{h.c.}(-\vec{k}) \,, \quad\quad\,\,
    \hat{\sigma}_{(0),\vec{k}} = \sum_{A=1}^2 \sigma_{(0),k}^{A} \hat{a}_{\vec{k}}^{A}  + \mathrm{h.c.}(-\vec{k}) \,, \\
    \hat{\gamma}_{\vec{k}}^\lambda &= \sum_{A=1}^2 \gamma_k^{\lambda,A} \hat{a}_{\vec{k}}^{\lambda,A}  + \mathrm{h.c.}(-\vec{k}) \,, \quad
    \hat{\sigma}_{(2),\vec{k}}^\lambda = \sum_{A=1}^2 \sigma_{(2),k}^{\lambda,A} \hat{a}_{\vec{k}}^{\lambda,A}  + \mathrm{h.c.}(-\vec{k}) \,,
\end{align}
with the following commutation relations:
\begin{equation}
    \left[\hat{a}^A_{\vec{k}}, \hat{a}^{B,\dagger}_{\vec{k}^\prime} \right] =  (2\pi)^3 \delta^{AB} \delta^{(3)}\left(\vec{k}-\vec{k}^\prime\right) \,, \quad \text{and} \quad   
    \left[\hat{a}^{\lambda,A}_{\vec{k}}, \hat{a}^{\lambda^\prime,B,\dagger}_{\vec{k}^\prime} \right] =  (2\pi)^3 \delta^{AB} \delta^{\lambda\lambda^\prime} \delta^{(3)}\left(\vec{k}-\vec{k}^\prime\right) \,.
\end{equation}

By solving the coupled linear equations of motions for the mode functions $(\zeta_k^{A}, \sigma_{(0),k}^{A})$ and  $(\gamma_k^{\lambda,A},\sigma_{(2),k}^{\lambda,A})$, one can compute all relevant power spectra.
For example, the tensor power spectrum at the end of inflation simply reads:
\begin{equation}
    P_\gamma(k)=\underset{ \tau \rightarrow 0}{\mathrm{lim}} \sum_{\lambda=+,\times}\sum_{A=1}^2 \left|\gamma^{\lambda,A}_ k(\tau)\right|^2=P_\gamma^{(1)}(k)+P_\gamma^{(2)}(k) \,,
\end{equation}
from which it is clear that it is made of two independent contributions. It is in general difficult to solve the fully coupled equations, but the work in \cite{Bordin:2018pca} showed how it is possible to estimate $\mathcal{P}_\gamma=k^3/(2\pi^2)P_\gamma(k)$ for a small helicity-2 sound speed:
\begin{equation}
\label{eq: spin2 non-perturbative tensor PS}
    \mathcal{P}_\gamma \underset{c_2 \ll 1}{\simeq} \mathcal{P}_\gamma^0 \times \frac{1}{2 c_2^{2\nu}} \left(\frac{6\rho H}{m^2+2\rho^2}\right)^2 \,, \quad \text{with} \quad \nu=\sqrt{\frac{9}{4}-\frac{m^2}{H^2}} \quad \text{and} \quad \mathcal{P}_\gamma^0=\frac{2H^2}{\pi^2 \Mp^2} 
\end{equation}
where $\mathcal{P}^0_{\gamma}$ is the usual vacuum tensor power spectrum.
The total tensor power spectrum in this model can be much  larger than in usual single-field models of inflation. At large scales it is naturally constrained by the upper bound on the tensor-to-scalar ratio $r$. However, the presence of the coupling constant $\rho$ and of the helicity-2 sound speed $c_2$ in the expression  for the two-point-function makes it possible to implement a blue-tilted spectrum, for example by means of a scale-dependent $c_2(k)$ \cite{Iacconi:2019vgc,Iacconi:2020yxn}.

\subsection{Primordial STT Bispectrum}

We consider the interaction of one scalar mode with two tensor modes, which is possible, e.g., through the interaction $\mathcal{L}^{(3)}_\sigma$ of Eq.~\eqref{eq: Lsigma}.
In the soft scalar limit, we expect the interaction to lead to gravitational waves anisotropies, with an amplitude set by the STT bispectrum. The interaction Hamiltonian we shall need  (this under the assumption that quadratic interactions have been taken fully into account in the equations of motion for the mode functions\footnote{We will not provide here the explicit solution for the mode functions in the non-perturbative regime, i.e. when the $\pi-\sigma_0$ and $\gamma-\sigma_2$ couplings cannot be considered small. See \cite{Bordin:2018pca} for more on the non-perturbative regime.}) is then:
\begin{equation}
    \int \dd t H_\mathrm{int}^{\sigma^3} = 3 \mu \int \dd \tau  a^4(\tau) \int\frac{\dd^3 \vec{k} \dd^3 \vec{q}}{(2\pi)^6} \sum_{\lambda,\lambda^\prime}
    \mathcal{A}^{\lambda\lambda^\prime}\cdot 
    \sigma_{(2),\vec{k}}^{\lambda}(\tau) \sigma_{(2),\vec{q}}^{\lambda^\prime}(\tau)
    \sigma_{(0),-\vec{k}-\vec{q}}(\tau)\,,
\end{equation}
where $\mathcal{A}^{\lambda\lambda^\prime}$ is a simple product of polarization tensors (see \cite{Dimastrogiovanni:2021mfs}).
\begin{figure}
    \centering
    \begin{subfigure}{0.48\textwidth}
        \centering
        \includegraphics[width=1.\linewidth]{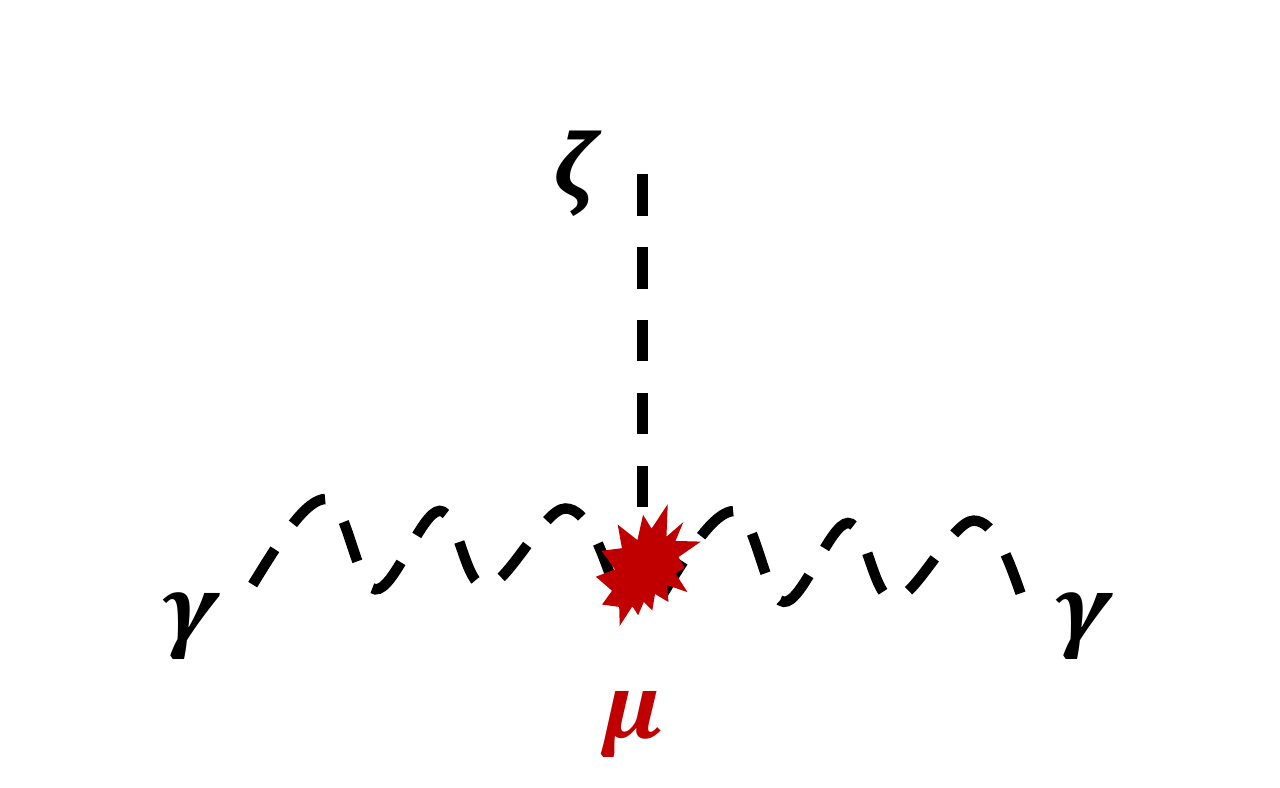}
        \caption{Mixed scalar-tensor-tensor bispectrum.}
    \end{subfigure}%
    \hfill
    \begin{subfigure}{0.48\textwidth}
        \centering
        \includegraphics[width=1.\linewidth]{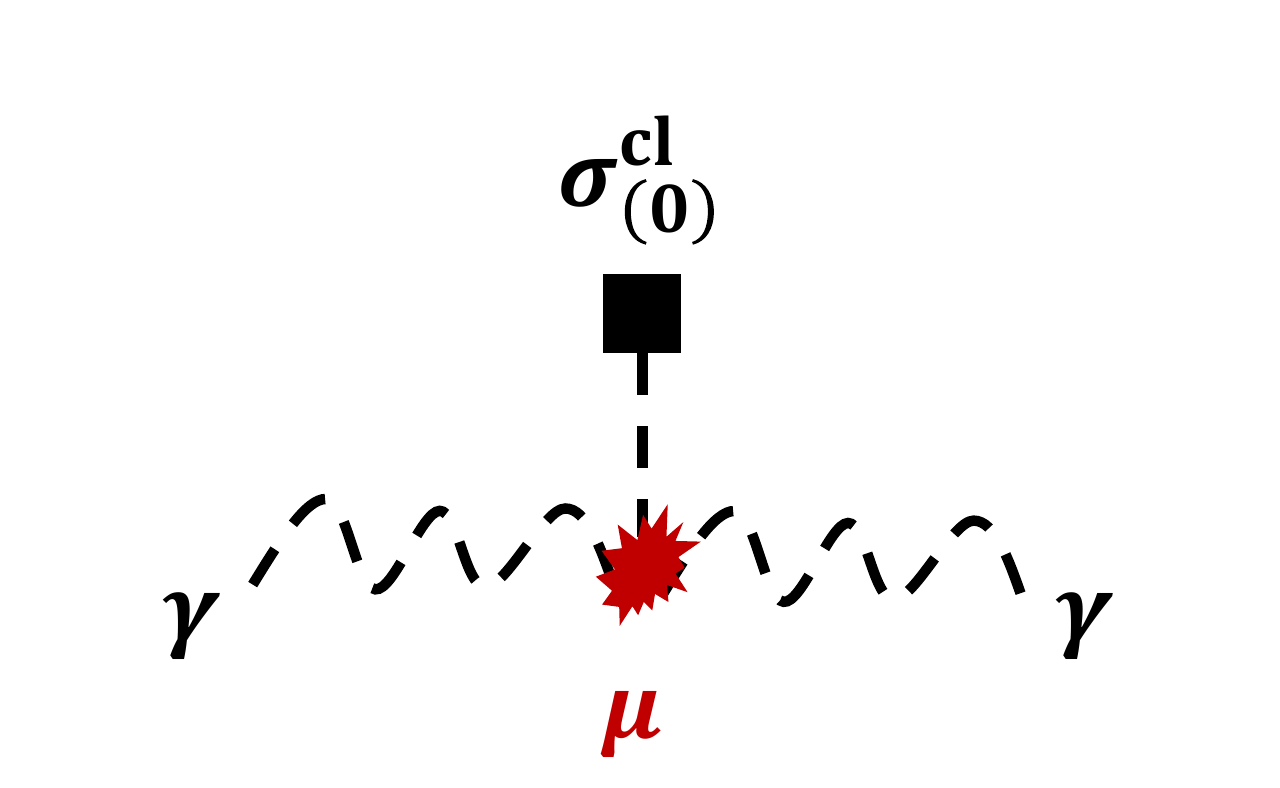}
        \caption{Tensor two-point function in the presence of a classical scalar source.}
    \end{subfigure}
    \caption{Relevant Feynman diagrams for the primordial anisotropies in the spin-2 case. Dashed lines represent mixed propagators of the $\zeta-\sigma_{(0)}$ type (straight lines), and the $\gamma-\sigma_{(2)}$ type (wavy lines). The square at the top of the dashed line represents, in contradistinction to a propagator connecting to an external leg, a classical background $\sigma_{(0)}^\mathrm{cl}$.}
    \label{fig: spin2 - non-pert}
\end{figure}
The STT bispectrum, whose diagram is in the left panel of Fig.~\ref{fig: spin2 - non-pert}, reads:
\begin{align}
\label{eq: first step spin2 BS}
    \sum_{\lambda,\lambda^\prime}\Braket{\hat{\gamma}^\lambda_{\vec{k}_1}\hat{\gamma}^{\lambda^\prime}_{\vec{k}_2}\hat{\zeta}_{\vec{k}_3}}^{\sigma^3}%(\tau_0)
    &=
    (2\pi)^3 \delta^{(3)}\left(\vec{k}_1+\vec{k}_2+\vec{k}_3\right)
    B^{\gamma\gamma\zeta}(k_1,k_2,k_3)%; \tau_0)
    \,,     \quad\text{ with } \\
    B^{\gamma\gamma\zeta}(k_1,k_2,k_3)%;\tau_0) 
    %&\underset{k_3 \ll k_1, k_2}{=} 
    =& 
    -12 \mu \mathrm{Im} \left\{  \sum_{\lambda,\lambda^\prime}
    \sum_{A,B,C}\mathcal{A}^{\lambda\lambda^\prime} 
    \gamma^{\lambda,A}_{k_1}
    \gamma^{\lambda^\prime,B}_{k_2}
    \zeta_{k_3}^C
    \int_{-\infty^+}^{0} \dd \tau a^4(\tau) 
    \sigma_{(2),k_1}^{\lambda,A,*}(\tau) \sigma_{(2),k_2}^{\lambda^\prime,B,*}(\tau)
    \sigma_{(0),k_3}^{C,*}(\tau) 
    \right\}
    \,, \nonumber
\end{align}

where we have taken into account the fact that quantum operators have, in light of the couplings in the quadratic theory, non-trivial cross-correlations: $\braket{0|\hat{\zeta}_{\vec{k}}\hat{\sigma}_{(0),\vec{k}^\prime}|0}=(2\pi)^3 \delta^{(3)}(\vec{k}+\vec{k}^\prime)\sum_A \zeta_k^A \sigma_{(0),k}^{A,*}$.
In the squeezed configuration, $k_3 = k_L \ll k_S = k_1 \simeq k_2$, one may factor the long scalar mode $\sigma_{(0),k_L}(\tau)$ out of the time integral. Indeed, the integral is dominated by conformal times close to $-\tau_S \sim 1/k_S$, at which the long mode has already reached its super-horizon value.
Note that, strictly speaking, if the extra spin-2 field has a non-zero mass, the mode function $\sigma_{(0)}$ does not become exactly constant on superhorizon scales.
However, when $\sigma_{(0)}$ is light, which we assume in the following\footnote{If $\sigma$ is heavy, $m_{\sigma}\sim H$, the (squeezed limit of the)  bispectrum will be suppressed. In the regime where the quadratic interactions are taken into account perturbatively, the power-law suppression of the squeezed bispectrum is well-known, $(k_L/k_S)^{3-2\nu}$.

}, one can neglect the logarithmic time-dependence on super-horizon scales and approximate the value of $\sigma_{(0)}$ at the time $\tau_S$ of horizon exit for the small scales under consideration with the one at the end of inflation at which the bispectrum is evaluated.
In the squeezed limit, the polarisation tensors simplify to  $\mathcal{A}^{\lambda\lambda^\prime} \simeq  \mathcal{A}(\hat{k}_L,\hat{k}_S) \delta^{\lambda\lambda^\prime}$ and
the STT bispectrum is:
\begin{align}
\label{eq: last step spin2 BS}
    B^{\gamma\gamma\zeta}(k_S,k_S,k_L) \underset{k_L \ll k_S}{=} &- P_{\zeta\sigma_{(0)}}(k_L)
    \times 12 \mu
    \times
    \mathcal{A}(\hat{k}_L,\hat{k}_S) \\
    & \times\mathrm{Im}
    \left\{
    \sum_{\lambda}
    \sum_{A,B}
    \gamma^{\lambda,A}_{k_S}
    \gamma^{\lambda,B}_{k_S}
    \int_{-\infty^+}^{0} \dd \tau a^4(\tau) 
    \sigma_{(2),k_S}^{\lambda,A,*}(\tau) \sigma_{(2),k_S}^{\lambda,B,*}(\tau) \right\}\,, \nonumber
\end{align}
where we have used the definition of the cross power-spectrum:
\begin{equation}
P_{\zeta\sigma_{(0)}} = \underset{\tau \rightarrow 0}{\mathrm{lim}} \sum_A \zeta^A_k(\tau)      \sigma_{(0),k}^{A,*}(\tau)\,.
\end{equation}

\subsection{GW Anisotropy}
Equipped with the knowledge of the same interaction, we now study its effect on the two-point function in the limit where the long scalar mode acts as a classical background, $
    \hat{\sigma}_{(0)} \longrightarrow \sigma_{(0)}^\mathrm{cl} $.

This amounts to the induced anisotropy of the form (see the right panel of Fig.~\ref{fig: spin2 - non-pert} for the corresponding diagram):
\begin{align}
    \sum_{\lambda,\lambda^\prime}\left.\Braket{\hat{\gamma}^\lambda_{\vec{k}_1}\hat{\gamma}^{\lambda^\prime}_{\vec{k}_2}}\right|_{\sigma_{(0)}^\mathrm{cl}}     =& 
    -12 \mu \mathrm{Im}\left\{   \sum_{\lambda,\lambda^\prime}
    \sum_{A,B}
    \gamma^{\lambda,A}_{k_1}
    \gamma^{\lambda^\prime,B}_{k_2}
    \times
    \mathcal{A}^{\lambda\lambda^\prime} \right. \\
    & \left.\times
    \int_{-\infty^+}^{0} \dd \tau a^4(\tau) 
    \sigma_{(2),k_1}^{\lambda,A,*}(\tau) \sigma_{(2),k_2}^{\lambda^\prime,B,*}(\tau)
    \sigma_{(0),\vec{k}_1+\vec{k}_2}^\mathrm{cl}(\tau)
    \right\}
    \,. \nonumber
\end{align}
Focusing on the momentum configuration  $|\vec{k}_1 + \vec{k}_2|\ll k_1, k_2$, one may simplify this expression further, similarly to what is done for the bispectrum in the squeezed limit. The quantity  $\sigma_{(0),\vec{k}_1+\vec{k}_2}^\mathrm{cl}$ can be taken outside of the integral and evaluated at the final time,  if, as we assume here, the spin-2 field is sufficiently light. The resulting expression can be written, using  Eq.~\eqref{eq: last step spin2 BS}, as: 
\begin{align}
\label{eq316}
    \sum_{\lambda,\lambda^\prime}\left.\Braket{\hat{\gamma}^\lambda_{\vec{k}_1}\hat{\gamma}^{\lambda^\prime}_{\vec{k}_2}}\right|_{\sigma_{(0)}^\mathrm{cl}} \underset{|\vec{k}_1 + \vec{k}_2| \ll k_1,k_2}{=} &
    \frac{\sigma_{(0),\vec{k}_1+\vec{k}_2}^\mathrm{cl}}{P_{\zeta \sigma_{(0)}}(|\vec{k}_1 + \vec{k}_2|)} B^{\gamma\gamma\zeta}(k_1,k_2,|\vec{k}_1 + \vec{k}_2|) \,.
\end{align}
The latter can be readily written as 

\begin{equation}
\label{eq: spin2 final anisotropies}
    \sum_{\lambda,\lambda^\prime}\left.\Braket{\hat{\gamma}^\lambda_{\vec{k}_1}\hat{\gamma}^{\lambda^\prime}_{\vec{k}_2}}\right|_{\sigma_{(0)}^\mathrm{cl}} \underset{|\vec{k}_1 + \vec{k}_2| \ll k_1,k_2}{=} \int \dd^3 \vec{q} \, \delta^{(3)} (\vec{q}+\vec{k}_1+\vec{k}_2)     \frac{B^{\gamma\gamma\zeta}(k_1,k_2,q)}{P_{\zeta \sigma_{(0)}}(q)}  \sigma_{(0),-\vec{q}}^\mathrm{cl}  \,,
\end{equation}
which is the STT counterpart of the (TTT)  heuristic formula we reported in Eq.~(\ref{q1}). We have seen so far then how Eq.~(\ref{q1}) may be arrived at via the application of the in-in formalism in the context of minimal single-field slow-roll inflation and also in the case of an extra spin-2 fields directly coupled to the inflaton. Before moving on to another application, we give here the relation between bispectrum and induced GW anisotropy in the small-quadratic-mixing regime of the spin-2 model.

\subsection{Explicit computation in the case of small mixing}
\label{sec: perturbative spin2}
\begin{figure}
    \centering
    \begin{subfigure}{0.48\textwidth}
        \centering
        \includegraphics[width=1.\linewidth]{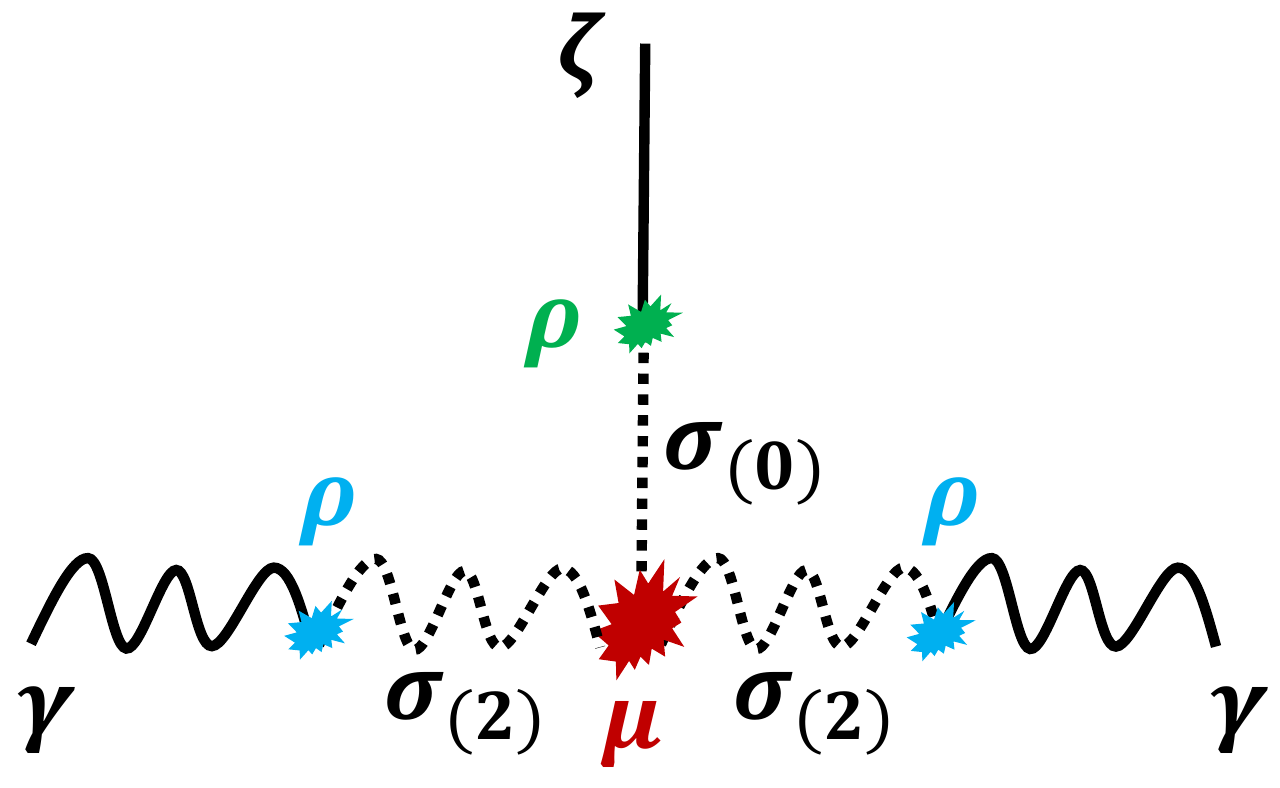}
        \caption{Mixed scalar-tensor-tensor bispectrum.}
    \end{subfigure}%
    \hfill
    \begin{subfigure}{0.48\textwidth}
        \centering
        \includegraphics[width=1.\linewidth]{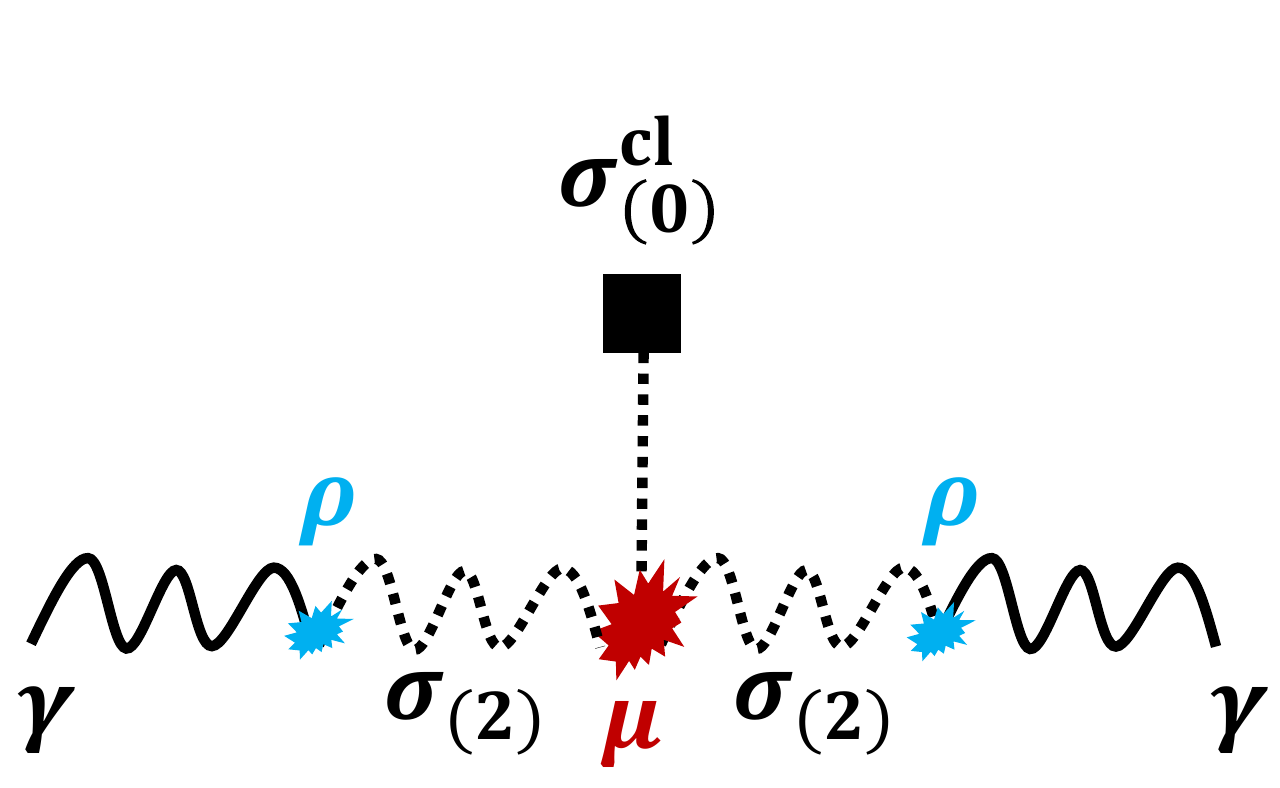}
        \caption{Tensor two-point function in the presence of a classical scalar source.}
    \end{subfigure}
    \caption{Relevant diagrams for the primordial anisotropies in the spin-2 model in the small mixing regime.
        Continuous lines represent propagators of $\zeta$ (straight) and $\gamma$ (wavy), while dotted lines represent propagators of $\sigma_{(0)}$ (straight) and $\sigma_{(2)}$ (wavy). The  square represents the classical source $\sigma_{(0)}^\mathrm{cl}$.}
    \label{fig: spin2 - pert}
\end{figure}

In the regime of a small quadratic mixing, $\mathcal{L}^{(2)}_\mathrm{mix}$ can be treated as a perturbation on top of the free field Lagrangian and incorporated in the interaction Hamiltonian. One can  proceed by quantising each field independently, recovering the usual mode functions for decoupled massless modes ($\zeta$ and $\gamma$) and light helicity modes ($\sigma_{(0),(2)}$, respectively with a speed of sound $c_{0,2}$) in de Sitter spacetime. In this framework, computing the mixed bispectrum requires the introduction of four vertices: one scalar 2-vertex, two tensor 2-vertices, and one mixed 3-vertex, as represented\footnote{ To the reader familiar with the subject, both this notation and nomenclature will be reminiscent of quasi-single-field (QSF) inflation \cite{Chen:2009we}. Indeed, the coupling in the tensor sector between $\gamma$ and $\sigma_{(2)}$ modes is completely analogous to the coupling in the scalar sector of QSF inflation, and so is the $\sigma_{(2)}$ cubic self-interaction.} in the left panel of Fig.~\ref{fig: spin2 - pert}.
One first finds:
\begin{align}
\label{eq317}
\Braket{\hat{\gamma}^\lambda_{\vec{k}_1}\hat{\gamma}^{\lambda^\prime}_{\vec{k}_2}\hat{\zeta}_{\vec{k}_3}}= & 
    \int_{-\infty}^0 \dd \tau_1 a(\tau_1) 
    \int_{-\infty}^{\tau_1} \dd \tau_2 a(\tau_2)
    \int_{-\infty}^{\tau_2} \dd \tau_3 a(\tau_3) 
    \int_{-\infty}^{\tau_3} \dd \tau_4 a(\tau_4) \\
    & \times \Braket{0|
    \left[\hat{H}_\mathrm{int}(\tau_4), 
    \left[\hat{H}_\mathrm{int}(\tau_3),
    \left[\hat{H}_\mathrm{int}(\tau_2),
    \left[\hat{H}_\mathrm{int}(\tau_1), 
    \hat{\gamma}^\lambda_{\vec{k}_1}\hat{\gamma}^{\lambda^\prime}_{\vec{k}_2}\hat{\zeta}_{\vec{k}_3}
    \right]\right]\right]\right]|0} \,, \nonumber
\end{align}
with
\begin{equation}
    H_\mathrm{int}=H_\mathrm{int}^{\zeta\sigma_{(0)}}+H_\mathrm{int}^{\gamma\sigma_{(2)}}+H_\mathrm{int}^{\sigma_{(0)}\sigma_{(2)}\sigma_{(2)}} \,.
\end{equation}

Note that the size of the contribution in Eq.(\ref{eq317}) depends on which interaction Hamiltonian takes the role of $\hat{H}_\mathrm{int}(\tau_4),...,\hat{H}_\mathrm{int}(\tau_1)$. For example,
given that in the small-mixing regime a given field can only be contracted with another one of the same type, the cubic Hamiltonian cannot feature as $\hat{H}_\mathrm{int}(\tau_1)$ in Eq.(\ref{eq317}).
If it did, the commutator would simply vanish as $[\hat{H}_\mathrm{int}^{\sigma_{(0)}\sigma_{(2)}\sigma_{(2)}} , \hat{\gamma}\hat{\gamma}\hat{\zeta}]=0$.
Enforcing the squeezed configuration hierarchy, $k_3=k_L \ll k_1 \simeq k_2$, will also further reduce the permutations that give rise to the leading contributions.
Indeed, if one places the interaction involving only soft momenta in the innermost time integral in $\dd\tau_4$, the useful domain of integration of this first integral begins as $\tau_L\sim-1/k_L$ (i.e. the horizon of the largest $k$ mode in that given time variable) whilst the rest of the time integrals will, for the same reason, start out at $\tau_S\sim-1/k_S$.
If instead we consider any other interaction in the innermost integral, the integration interval will necessarily start out at $\tau_S\sim-1/k_S$, and so will all other integrals in light of the $\tau_1\geq\tau_2\geq\tau_3$ hierarchy.
It follows that only the permutations that place the interaction with only long-modes in the innermost integral will give rise to the leading contribution, by virtue of the fact its effective integration domain over time is parametrically larger\footnote{We have verified this both analytically (for the massless case) and numerically (for generic $\nu$).}. 
One can also show, again only by enforcing the squeezed limit hierarchy, that the (initially nested) integral over the soft modes along $\tau_4$ can be factored out, its upper limit of integration being very-well approximated by zero \cite{Chen:2009zp}.
One is  then left with only two permutations (depending on whether the cubic interaction  features as $\hat{H}_\mathrm{int}(\tau_2)$ or $\hat{H}_\mathrm{int}(\tau_3)$) giving the leading contributions: 
\begin{align}
\Braket{\hat{\gamma}^\lambda_{\vec{k}_1}\hat{\gamma}^{\lambda^\prime}_{\vec{k}_2}\hat{\zeta}_{\vec{k}_3}} \underset{k_3 \ll k_1,k_2}{=} & 
    \int_{-\infty}^0 \dd \tau_1 a(\tau_1) 
    \int_{-\infty}^{\tau_1} \dd \tau_2 a(\tau_2)
    \int_{-\infty}^{\tau_2} \dd \tau_3 a(\tau_3) 
    \int_{-\infty}^{0} \dd \tau_4 a(\tau_4) \\
    & \times \Braket{0|
    \left[\hat{H}_\mathrm{int}^{\zeta\sigma_{(0)}}(\tau_4), 
    \left[\hat{H}_\mathrm{int}^{\sigma_{(0)}\sigma_{(2)}\sigma_{(2)}}(\tau_3),
    \left[\hat{H}_\mathrm{int}^{\gamma\sigma_{(2)}}(\tau_2),
    \left[\hat{H}_\mathrm{int}^{\gamma\sigma_{(2)}}(\tau_1),
    \hat{\gamma}^\lambda_{\vec{k}_1}\hat{\gamma}^{\lambda^\prime}_{\vec{k}_2}\hat{\zeta}_{\vec{k}_3}
    \right]\right]\right]\right]|0}  \nonumber \\
    &+ (\tau_2 \leftrightarrow \tau_3)\,. \nonumber
\end{align}
Once the operators at $\tau_4$ are contracted, one obtains a factor of
\begin{equation}
    \zeta_{k_3}(\tau_4)\sigma_{(0),k_3}(\tau_4)\zeta_{k_3}^*\sigma_{(0),k_3}^*(\tau_3) - \text{c.c.}
\end{equation}
We also note that, given that the useful domain over the variable $\tau_3$ starts at $\tau_S=-1/k_S$, the function  $\sigma_{(0),k_3}^*(\tau_3)$ can be factored out of the integral, to give
\begin{align}
\label{eq: spin2 BS perturbative}
    \Braket{\hat{\gamma}^\lambda_{\vec{k}_1}\hat{\gamma}^{\lambda^\prime}_{\vec{k}_2}\hat{\zeta}_{\vec{k}_3}} \underset{k_3 \ll k_1,k_2}{=} & 
    -\sqrt{\frac{2}{3}} \frac{\rho \Mp}{H} k_3^2 
    \mathrm{Im}\left\{\zeta_{k_3}^*\sigma_{(0),k_3}^*
    \int^0_{-\infty^-} \dd\tau_4 a^2(\tau_4) \zeta_{k_3}(\tau_4)\sigma_{(0),k_3}(\tau_4)
    \right\} \\
    & \times
    \int_{-\infty}^0 \dd \tau_1 a(\tau_1) 
    \int_{-\infty}^{\tau_1} \dd \tau_2 a(\tau_2)
    \int_{-\infty}^{\tau_2} \dd \tau_3 a^4(\tau_3)
    \int \frac{\dd^3 \vec{k}\dd^3 \vec{q}}{(2\pi)^6} \nonumber \\
    & \times (2\pi)^3 \delta^{(3)}\left(\vec{k}_3-\vec{k}-\vec{q}\right)  3 \mu \sum_{\tilde{\lambda}, \tilde{\lambda}^\prime}
   \mathcal{A}^{\tilde{\lambda}\tilde{\lambda}^\prime}
    \nonumber \\
    & \times \Braket{0|
    \left[\hat{\sigma}_{(2),\vec{k}}^{\tilde{\lambda}}(\tau_3)\hat{\sigma}_{(2),\vec{q}}^{{\tilde{\lambda}}^\prime}(\tau_3),
    \left[\hat{H}_\mathrm{int}^{\gamma\sigma_{(2)}}(\tau_2),
    \left[\hat{H}_\mathrm{int}^{\gamma\sigma_{(2)}}(\tau_1),
    \hat{\gamma}^\lambda_{\vec{k}_1}\hat{\gamma}^{\lambda^\prime}_{\vec{k}_2}
    \right]\right]\right]|0}  \nonumber \\
    &+ (\tau_2 \leftrightarrow \tau_3)\,, \nonumber
\end{align}
where in the first line we recognize the exact expression for the cross power spectrum $P_{\zeta\sigma_{0}}(k_3)$ (at first non-vanishing order in $\rho/H$) found via the in-in formalism:
\begin{equation}
    \label{eq: cross power spectrum spin2}
    P_{\zeta\sigma_{(0)}}(k) =-\sqrt{\frac{2}{3}} \frac{\rho \Mp}{H} k^2 \mathrm{Im}\left\{\zeta_k^*\sigma_{(0),k}^*
    \int^0_{-\infty^-} \dd\tau_1 a^2(\tau_1) \zeta_k(\tau_1)\sigma_{(0),k}(\tau_1)
    \right\} \,.
\end{equation}
This cross power spectrum may be evaluated exactly.
In the massless case $\nu=3/2$, we find  $P_{\zeta\sigma_{0}}(k) = [(2\pi^2)/k^3]\mathcal{P}_{\zeta\sigma_{0}}$ with:
\begin{equation}
    \label{eq: cross power spectrum spin2 explicit}
    \mathcal{P}_{\zeta\sigma_{(0)}} \underset{\nu=3/2}{=} \Mp \sqrt{\frac{3}{8}} \frac{\rho}{H} \mathcal{P}_\zeta\,, \quad \text{and} \quad\mathcal{P}_\zeta=\frac{H^2}{8\pi^2\epsilon\Mp^2} \,.
\end{equation}
The result of the explicit calculation of the mixed bispectrum in Eq.~\eqref{eq: spin2 BS perturbative} can be found in Ref.~\cite{Dimastrogiovanni:2021mfs}, where it is shown that:
\begin{align}
\label{eq: spin2 BS perturbative explicit}
    \sum_{\lambda,\lambda^\prime}\Braket{\hat{\gamma}^\lambda_{\vec{k}_1}\hat{\gamma}^{\lambda^\prime}_{\vec{k}_2}\hat{\zeta}_{\vec{k}_3}} =  &\, (2\pi)^3 \delta^{(3)}\left(\vec{k}_1+\vec{k}_2+\vec{k}_3\right)
    B^{\gamma\gamma\zeta}(k_1,k_2,k_3)
    \,,     \quad\text{ with } \nonumber \\
    B^{\gamma\gamma\zeta}(k_S,k_S,k_L)
    \underset{k_L\ll k_S}{=} &  -  2\pi^2 \frac{H^3}{\epsilon \Mp^3}
    \frac{\mu}{H}\left(\frac{\rho}{H}\right)^3
    \frac{2^\nu}{k_S^{9/2-\nu}k_L^{3/2+\nu}}
    \mathcal{A}\left(\hat{k}_L,\hat{k}_S\right)
    \mathcal{I}(c_0,c_2,\nu)   \,. 
\end{align}
We refer the reader to \cite{Dimastrogiovanni:2021mfs} for a semi-analytical description of the function $\mathcal{I}(c_0,c_2,\nu)$.

On the other hand, the tensor two-point function in the presence of the classical source $\sigma_{(0)}^\mathrm{cl}$ (see, in Fig.~\ref{fig: spin2 - pert}, the right diagram corresponding to the small-mixing regime) reads:

\begin{align}
\label{onet}
    \sum_{\lambda,\lambda^\prime}\left.\Braket{\hat{\gamma}^\lambda_{\vec{k}_1}\hat{\gamma}^{\lambda^\prime}_{\vec{k}_2}}\right|_{\sigma_{(0)}^\mathrm{cl}} = &
    \sum_{\lambda,\lambda^\prime}
    \int_{-\infty}^0 \dd \tau_1 a(\tau_1) 
    \int_{-\infty}^{\tau_1} \dd \tau_2 a(\tau_2)
    \int_{-\infty}^{\tau_2} \dd \tau_3 a^4(\tau_3)
    \int \frac{\dd^3 \vec{k}\dd^3 \vec{q}}{(2\pi)^6}
    \sigma_{(0),-\vec{k}-\vec{q}}^\mathrm{cl}
    \nonumber \\
    & \times % (2\pi)^3 \delta^{(3)}\left(\vec{k}_3-\vec{k}-\vec{q}\right)
    3 \mu \sum_{\tilde{\lambda}, {\tilde{\lambda}}^\prime}
    \mathcal{A}^{\tilde{\lambda}\tilde{\lambda}^\prime}
    \nonumber \\
    & \times \Braket{0|
    \left[\hat{\sigma}_{(2),\vec{k}}^{\tilde{\lambda}}(\tau_3)\hat{\sigma}_{(2),\vec{q}}^{{\tilde{\lambda}}^\prime}(\tau_3),
    \left[\hat{H}_\mathrm{int}^{\gamma\sigma_{(2)}}(\tau_2),
    \left[\hat{H}_\mathrm{int}^{\gamma\sigma_{(2)}}(\tau_1),
    \hat{\gamma}^\lambda_{\vec{k}_1}\hat{\gamma}^{\lambda^\prime}_{\vec{k}_2}
    \right]\right]\right]|0}  \nonumber \\
    &+ (\tau_2 \leftrightarrow \tau_3)\,.
\end{align}

The two tensor modes at $\tau_3$ will be contracted with the remaining tensors such that, by momentum conservation, one finds $(\vec{k},\vec{q})\rightarrow(-\vec{k}_1,-\vec{k}_2)$, the other permutation being equivalent.
The classical source has momentum $\vec{k}_1+\vec{k}_2$ which in the quasi-diagonal configuration satisfies the inequality $|\vec{k}_1 + \vec{k}_2| \ll k_1,k_2$. One can show, in complete analogy with the bispectrum case, that  here too the (light) classical source is well-approximated, for the purpose of the time integrals, by its constant value outside its horizon and can therefore be factored out completely.  

One may then compare the resulting expression with the mixed bispectrum in the soft scalar limit, Eq.~\eqref{eq: spin2 BS perturbative}, obtaining: 
\begin{align}
    \sum_{\lambda,\lambda^\prime}\left.\Braket{\hat{\gamma}^\lambda_{\vec{k}_1}\hat{\gamma}^{\lambda^\prime}_{\vec{k}_2}}\right|_{\sigma_{(0)}^\mathrm{cl}} \underset{|\vec{k}_1 + \vec{k}_2| \ll k_1,k_2}{=} &
    \frac{\sigma_{(0),\vec{k}_1+\vec{k}_2}^\mathrm{cl}}{P_{\zeta \sigma_{(0)}}(|\vec{k}_1 + \vec{k}_2|)} B^{\gamma\gamma\zeta}(k_1,k_2,|\vec{k}_1 + \vec{k}_2|) \,,
\end{align}
where we have used the fact that one may identify and isolate the scalar cross power spectrum in the first line of Eq.~\eqref{eq: spin2 BS perturbative}. This large-scale modulation of the tensor two-point function gives rise to anisotropies of primordial origin in the SGWB, with a typical amplitude, at a scale $k$, of:
\begin{align}
    \sqrt{\braket{\delta_\mathrm{GW}^2(k,\hat{n})} } & \sim   
    \frac{B^{\gamma\gamma\zeta}(k,k,k_*\ll k)}{P_\gamma(k) P_{\zeta\sigma_{(0)}}(k_*)} \sqrt{\mathcal{P}_{\sigma_{(0)}}(k_*)} \,, \quad 
\end{align}
where $k_*$ can be taken, for example, at CMB scales\footnote{For completeness, we should quote the value of the almost scale-invariant power spectrum of the light extra scalar $\sigma_{(0)}$ at the end of inflation, $\mathcal{P}_{\sigma_{(0)}}=[k^3/(2\pi)^2]P_{\sigma_{(0)}}(k)$. In the perturbative treatment of a small quadratic mixing, it is, neglecting the slow super-horizon evolution, equal to:
$$\mathcal{P}_{\sigma_{(0)}}= \left(\frac{H}{2\pi}\right)^2 \times 2^{2\nu-3} \left(\frac{\Gamma(\nu)}{\Gamma(3/2)}\right)^2 \,.$$
}.

\section{Supersolid inflation}
\label{suso}

We are now going to consider an application to the case of supersolid inflation. Much like the spin-2 setup, this model can support (i) a blue GW spectrum and (ii) a non-trivial squeezed STT primordial non-Gaussianity. These are the prerequisites for having interesting, and possibly detectable at intermediate-small scales, GW anisotropies. We will briefly introduce the supersolid field content and refer the reader to the rich literature on the subject (see e.g. \cite{Bartolo:2015qvr,Ricciardone:2016lym,Celoria:2020diz,Celoria:2021cxq}) for more details. The specific model we shall consider is the one of \cite{Celoria:2020diz,Celoria:2021cxq}. 

Supersolid inflation belongs to a class of models that display a non-standard symmetry breaking pattern. A notable early example is
solid inflation \cite{Gruzinov:2004ty,Endlich:2012pz}: this is an interesting set-up showing how a successful acceleration mechanism may be obtained with a space-diffeomorphism breaking background, rather than the standard time-reparametrisation breaking (time-diffeomorphisms being non-linearly realised).
The supersolid model whose GW signatures we are after features the complete breaking of spacetime diffeomorphisms by the background configuration of (four) scalar fields $\varphi^A$.
Homogeneous background solutions exist in light of internal symmetries of the scalars \cite{Celoria:2020diz}. We refer the interested reader to \cite{Celoria:2017bbh} for a thorough classification of other interesting possibilities.

For our purposes we shall adopt the model in \cite{Celoria:2020diz} and it will suffice to mention here the presence of two propagating scalar degrees of freedom, sometimes termed  ``phonons''. 
The most immediate difference with respect to the spin-2 theory is the fact that the leading contribution to the GW spectrum here is the one non-linearly sourced  by scalars. We shall first illustrate the non-linear power spectrum result and then provide the STT calculation as well as the result for the induced anisotropy. The existing literature on this model includes a rough order-of-magnitude estimate of the amplitude of the one-loop tensor power spectrum \cite{Celoria:2020diz} whilst no calculation is present for the STT bispectrum. We provide here for the first time an exact calculation of both these observables as well as (also for the first time) the anisotropy induced in the tensor power spectrum by the STT three-point function.

We start by reviewing  scalar fluctuations and  considering their effect on the tensor power spectrum.

\subsection{Scalar fluctuations}

In flat gauge, the scalar fields $\varphi^A$  are written as:
\begin{equation}
    \varphi^0 = \bar{\varphi}(t) + \pi_0 \,, \quad \varphi^i = x^i + \partial^i \pi_L + \pi^i_T \,.
\end{equation}
The dynamics of the background field $\bar{\varphi}$ follows from the choice of an effective parameter $c_b^2$. In order to arrive at an almost scale-invariant spectrum of adiabatic fluctuations, it should be chosen to be close to either $0$ or $-1$ (see \cite{Celoria:2020diz}). 
The vector mode $\pi^i_T$, verifying the transverse condition $\partial_i \pi_T^i = 0$, decays rather quickly and it is therefore possible to neglect it in what follows.
In supersolid inflation then there are two propagating scalar degrees of freedom. We can group them together in the field-space vector $\pi^\alpha=(\pi_L,\pi_0)^\alpha$.
Their dynamics is regulated by the following quadratic action (written in conformal time):

\begin{equation}
    S_2 = \int \dd \tau \frac{\dd^3 \vec{k}}{(2\pi)^3} \left(
    \frac{1}{2} D_{\alpha\beta} \pi_{\vec{k}}^{\alpha\prime} \pi_{-\vec{k}}^{\beta\prime}
    + \Omega_{\alpha\beta} \pi_{\vec{k}}^{\alpha\prime}  \pi_{-\vec{k}}^\beta
    - \frac{1}{2}  M_{\alpha\beta} \pi_{\vec{k}}^\alpha \pi_{-\vec{k}}^\beta
    \right) \,.
\end{equation}
$D$ is a positive-definite diagonal matrix for the kinetic terms, $\Omega$ an anti-symmetric mixing matrix and $M$ a symmetric mass matrix also function of the wavenumber $k$.
They are all time-dependent.
The two Goldstone bosons $\pi_L, \pi_0$ define each a curvature perturbation as:
\begin{align}
    \zetan &= \frac{k^2}{3} \pi_L \,,  \\
    \Rpi &=  \frac{\mathcal{H}}{\bar{\varphi}^\prime} \pi_0\,.
\end{align}
The notation is motivated \cite{Celoria:2020diz} by the fact that $\zetan$ is defined as the curvature perturbation of hypersurfaces of constant number density $n$ of supersolid particles (rather than the total overdensity $\delta\rho$ for the usual $\zeta$), while $\Rpi$ is the curvature perturbation that is comoving with respect to the $\pi_0$ field only.  The curvature perturbation $\zeta$ can be related linearly to the previous quantities as:
\begin{equation}
    \zeta = \left(1-2 c_0^2 c_b^2 \right) \zetan - \frac{2c_0^2}{3 \bar{\varphi}^\prime} \pi_0^\prime \,,
\end{equation}
where $c_0^2$ is another background parameter of the EFT. 
It is possible to show  that it is $\zetan$ that
seeds the adiabatic, almost scale-invariant initial conditions for the anisotropies of the CMB, while $\Rpi$ only contributes a negligible amount of isocurvature modes~\cite{Celoria:2020diz}.

Because of the non-diagonal pieces of $\Omega$ and $M$, one may not assign a single fundamental quantum oscillator to each scalar: mixing needs to be taken into account. The quantization procedure is:
\begin{equation}
    \pi^\alpha_{\vec{k}} \rightarrow \hat{\pi}^\alpha_{\vec{k}} =  \sum_{A=1}^2 \pi_{A,k}^\alpha\hat{a}_{\vec{k}}^A + \pi_{A,k}^{\alpha*} \hat{a}^{A\dagger}_{-\vec{k}}\,\,, \quad \text{with} \quad \left[\hat{a}^A_{\vec{k}}, \hat{a}^{B,\dagger}_{\vec{k}^\prime} \right] =  (2\pi)^3 \delta^{AB} \delta^{(3)}\left(\vec{k}-\vec{k}^\prime\right) \,.
\end{equation}
Consistent initial conditions can be imposed on the mode functions $\pi_{A,k}^\alpha(\tau)$ by inspecting the sub-horizon limit of the quadratic action, from which one identifies the presence of two speeds of sound for scalar fluctuations, $c_{sA}$ with $A \in \{1,2\}$.

One can choose, without loss of generality, to consider the case $c_{s2} < c_{s1}$. In the following, we will focus on the possibility of the following hierarchy between the two sound speeds: $c_{s2} \ll c_{s1} \simeq 1$.\footnote{Henceforth, following the notations of \cite{Celoria:2020diz}, we further set the parameter $c_L^2=1/2$ as it was shown to be the value that is least restrictive on the other parameters.}

In \cite{Celoria:2020diz}, the coupled linear equations of motion for the scalar modes are solved perturbatively in a ``slow-roll" expansion in terms of $\epsilon, \eta$ and $\delta_b = 1+ c_b^2$ for the particular case $c_b^2 \simeq -1$ on which we shall focus from now on.

At zeroth order in generalized slow-roll one finds:
\begin{align}
    \hat{\zeta}_{n,\vec{k}}(\tau) &= \sqrt{P_0(k)} (-k\tau)^2\sum_{A=1}^2 C_L^A \sqrt{-k c_{sA} \tau} H_{5/2}^{(1)}(-k c_{s A} \tau) \hat{a}_{\vec{k}}^A + \, \mathrm{h.c.} \,, \\
    \hat{\mathcal{R}}_{\pi_0, \vec{k}}(\tau) &= \sqrt{P_0(k)} (-k\tau)\sum_{A=1}^2 C_0^A \sqrt{-k c_{sA} \tau} H_{3/2}^{(1)}(-k c_{s A} \tau) \hat{a}_{\vec{k}}^A + \, \mathrm{h.c.} \,,
\end{align}
with $P_0(k)$ the usual power spectrum of single-field slow-roll inflation:
\begin{equation}
    P_0(k)=\frac{2\pi^2}{k^3} \mathcal{P}_0 \,, \quad \text{with} \quad   \mathcal{P}_0 = \frac{H^2}{8 \pi^2 \epsilon \Mp^2}\; .
\end{equation}
The constant factors $C_{0,L}^A$ are determined from the Bunch-Davies vacuum conditions, and $H_{\nu}^{(1)}$ is the Hankel function of the first kind. At this order in slow-roll, the  scalar power spectra are indeed scale-invariant and read~\cite{Celoria:2020diz}: 
\begin{align}
    \mathcal{P}_{\zetan} &= \barP \mathcal{P}_0 \,, \quad\quad 
    \mathcal{P}_{\zetan \Rpi} = \frac{\beta}{c_{s2}^3} \mathcal{P}_0 \,, \quad\quad
    \mathcal{P}_{\Rpi} = \frac{\gamma}{c_{s2}^6} \mathcal{P}_0 \,, \\
    \text{ with} \quad  \barP & \sim 10  \quad \text{ and } \quad \beta \sim
    \gamma \sim 1 \nonumber \,,
\end{align}
where $\mathcal{P}_{\zetan \Rpi}$ is non-zero due to the correlation between the two scalar modes. Note that we have denoted $\mathcal{P}_{\zetan\zetan},\mathcal{P}_{\Rpi\Rpi}$ respectively as $\mathcal{P}_{\zetan},\mathcal{P}_{\Rpi}$ for simplicity.
The explicit expressions for $\barP, \beta, \gamma$ in terms of all the background parameters of the EFT can be found in Ref.~\cite{Celoria:2020diz}. For our purposes, it suffices here to know their typical order of magnitude and the scaling of the power spectra with respect to $c_{s2}$.
One should note the parametric enhancement, $1/c_{s2}^6 \gg 1$, for the power spectrum of $\Rpi$ compared to the one of $\zetan$ (and the one $1/c_{s2}^3 \gg 1 $ for the cross power spectrum).
At this stage, we stress it is important to characterise the statistics of $\Rpi$ on super-horizon scales even though, as already mentioned, only  $\zetan$ survives for reheating. The dynamics of $\Rpi$ is relevant because $\zetan$ and $\Rpi$ mix already at the linear level. Secondly, as we will see in the next section, $\Rpi$ non-linearly sources the production of tensor modes and therefore leaves an imprint on cosmological perturbations through this integrated-over-time effect.

Considering the next order in slow-roll makes the  deviations from scale-invariance in the power spectra manifest~\cite{Celoria:2020diz}:
\begin{align}
    \mathcal{P}_{\zetan}(k) &= \mathcal{P}_{\zetan}(k_*)\left( \frac{k}{k_*}\right)^{n_s^{(\mathrm{ad})}-1} \,, \quad\quad
    \mathcal{P}_{\Rpi}(k) = \sum_{(\alpha)\in\{\mathrm{ad},\mathrm{ad-en},\mathrm{en}\}} \mathcal{P}_{\Rpi}^{(\alpha)}(k_*) \left(\frac{k}{k_*} \right)^{n_s^{(\alpha)} -1} \,, 
\end{align}
with
\begin{align}
    n_s^{(\mathrm{ad})} &=  1+ \epsilon - \eta \,, \quad
    n_s^{(\mathrm{en})} = 1 + 6 \left(\delta_b - \epsilon\right) + \eta \,, \quad
    n_s^{(\mathrm{ad-en})} = \frac{n_s^{(\mathrm{ad})} + n_s^{(\mathrm{en})} }{2} \,. 
\end{align}
Given that $\zetan$-fluctuations are converted to the curvature perturbation $\zeta$ during reheating, the corresponding power spectrum is the one of primordial adiabatic fluctuations constrained by CMB observations.  The presence of another scalar degree of freedom, $\pi_0$, leads to a second mode with a different, independent, tilt: $n_s^{(\mathrm{en})}$, where the notation ``en" stands for entropic.
It is therefore possible to have $\Rpi$-fluctuations with a slightly blue tilt, so long as one considers the case \cite{Celoria:2020diz}
\begin{equation}
    \delta_b > \epsilon - \eta/6 \,.
\end{equation}
Given that $\eta$ is positive in order to have a red-tilted adiabatic spectrum, this condition is not very restrictive. We shall now turn to the observables associated with the tensor sector.

\subsection{Tensor power spectrum}

At the linear level, tensor modes of the metric display the exact same properties as in single-field slow-roll models (SFSR). Scalars can, of course, also source tensors at the non-linear level.
In SFSR the 1-loop contribution is suppressed by the standard $\left(H/\Mp\right)^2$ factor. In contradistinction to such case, in supersolid inflation the presence of the scalar mode $\Rpi$ can dramatically change the picture.
Furthermore, as we have seen, regions of the parameter space of the model support a blue tilt for $\mathcal{P}_{\Rpi}$, thus making the non-linear  contribution to the tensor power spectrum particularly relevant towards intermediate and small scales.

Let us proceed with the one-loop calculation with the in-in formalism. Note that an estimate for such quantity has been provided in \cite{Celoria:2020diz}. Later in this section we will calculate the one-loop STT bispectrum in the squeezed limit and its modulation effect on the tensor power spectrum. 

\bigskip

\begin{figure}
    \centering
    \begin{subfigure}{0.30\textwidth}
        \centering
        \includegraphics[width=1.\linewidth]{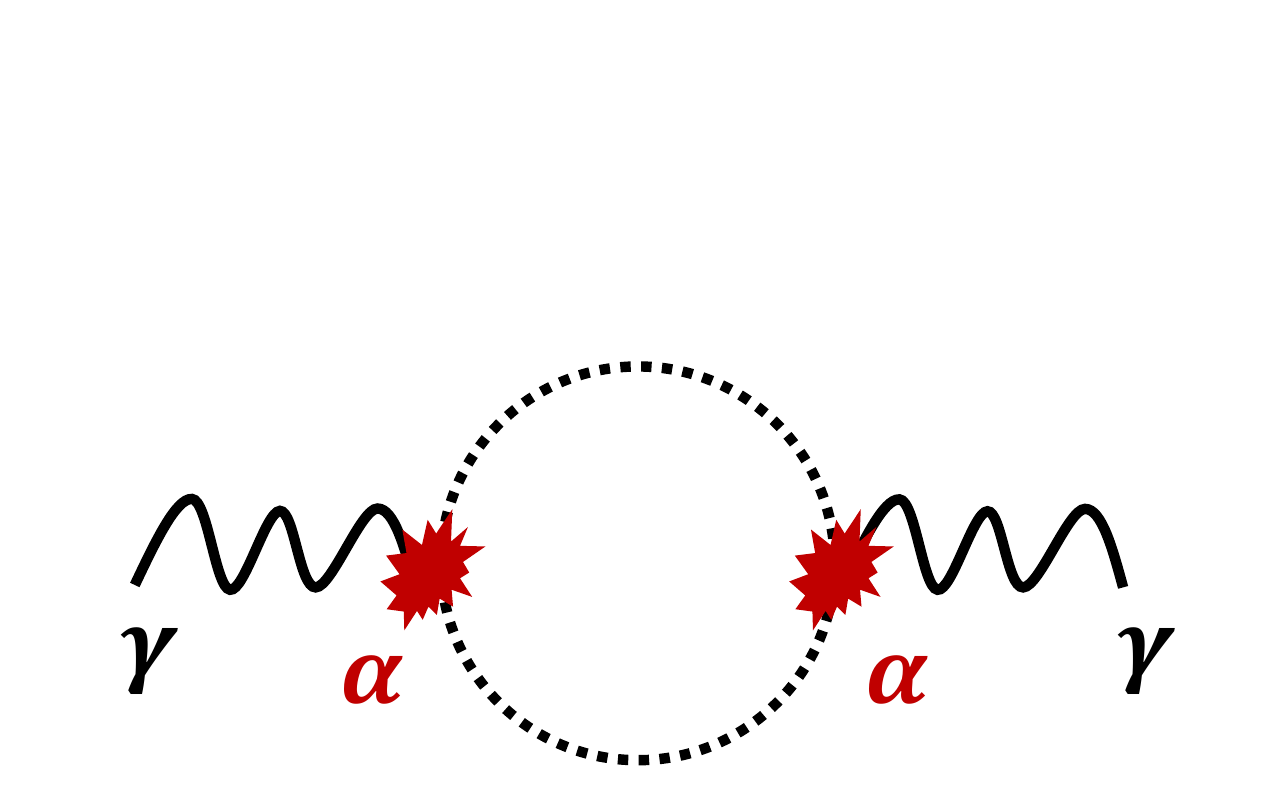}
        \caption{One-loop tensor power spectrum}
    \end{subfigure}%
    \hfill
    \begin{subfigure}{0.30\textwidth}
        \centering
        \includegraphics[width=1.\linewidth]{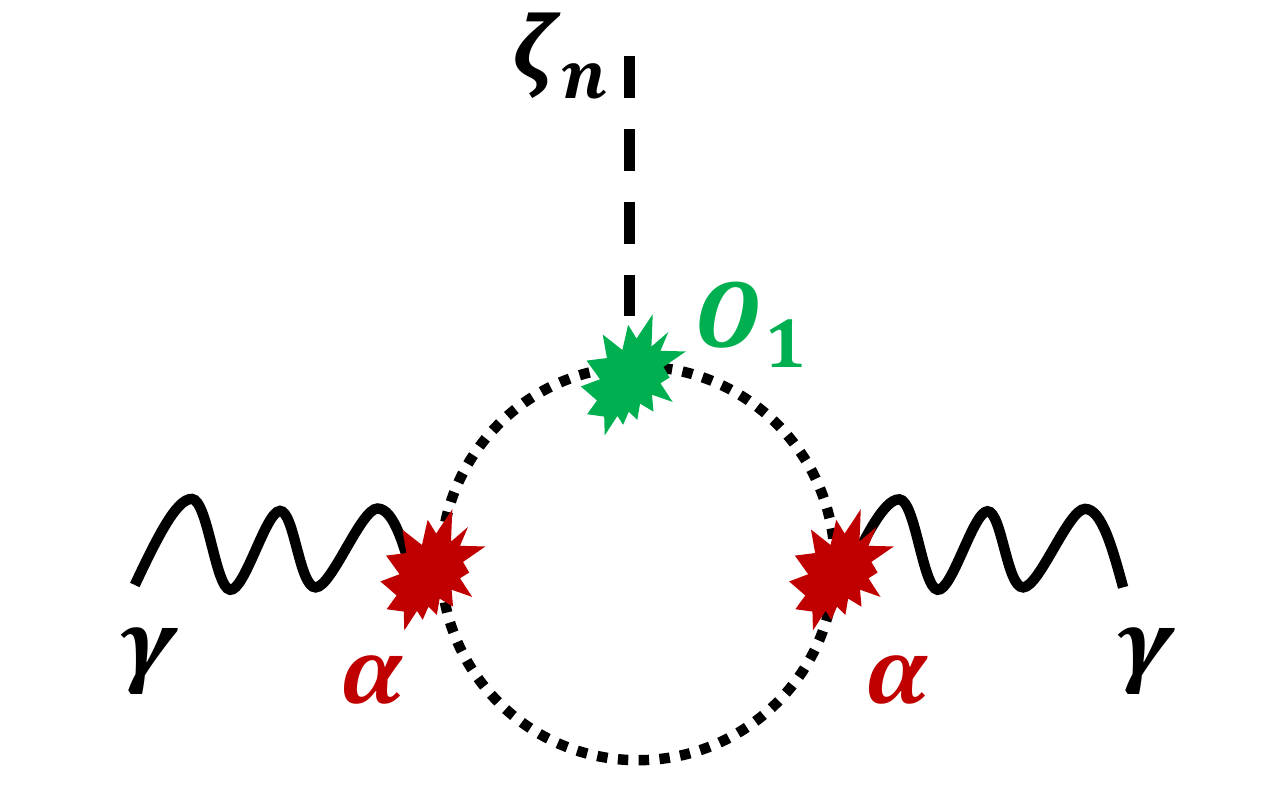}
        \caption{One-loop scalar-tensor-tensor bispectrum}
    \end{subfigure}
    \hfill
    \begin{subfigure}{0.30\textwidth}
        \centering
        \includegraphics[width=1.\linewidth]{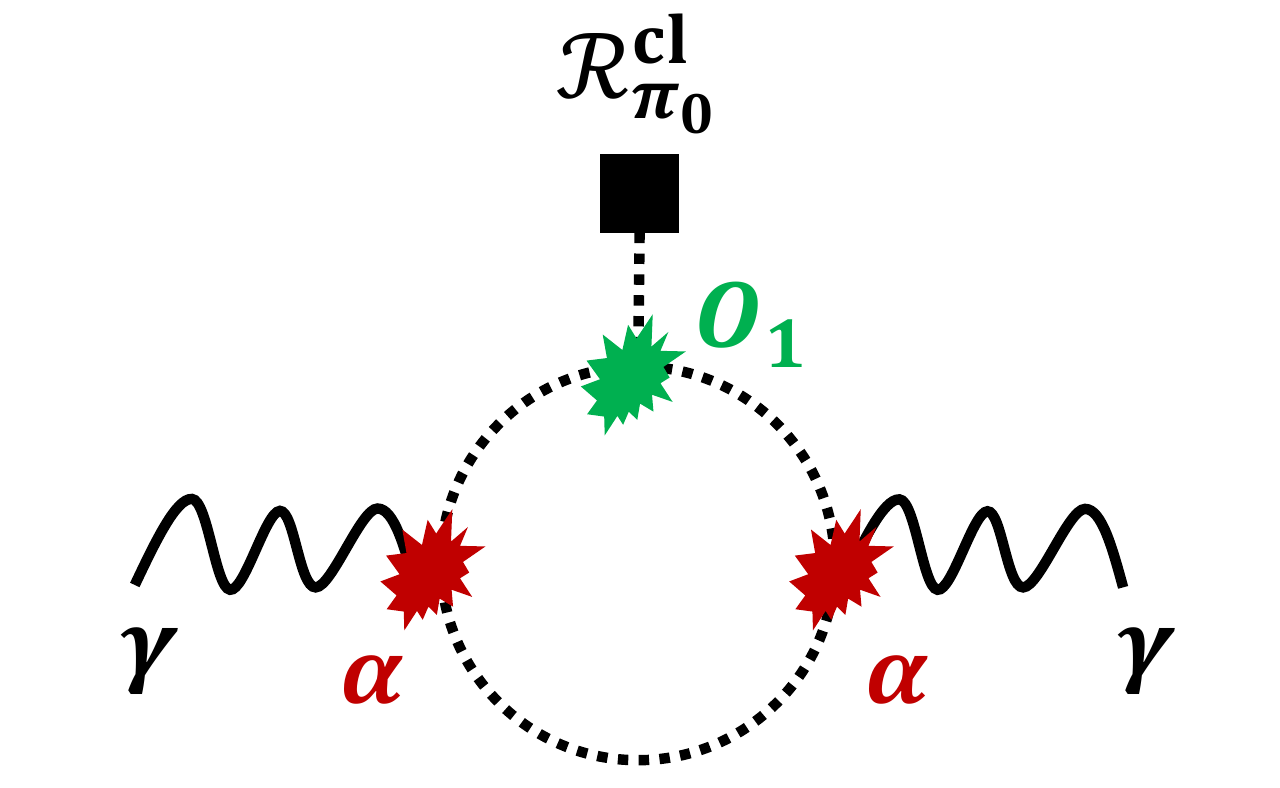}
        \caption{One-loop tensor two-point function in the presence of a classical scalar source.}
    \end{subfigure}
    \caption{Relevant diagrams for the primordial anisotropies in supersolid inflation. Wavy plain lines represent propagators of $\gamma$ while dotted lines represent propagators of $\Rpi$.
    The dashed line represents a mixed $(\zetan,\Rpi)$ propagator, and the dotted line ending with a square represents a classical random fluctuation $\Rpi^\mathrm{cl}$ rather than a propagator.}
    \label{fig: supersolid}
\end{figure}

We consider the gravitational cubic interaction of the scalars $\Rpi$ with the tensor $\gamma$, as:
\begin{equation}
\label{eq: TSS Hamiltonian supersolid}
    \int \dd t H_\mathrm{int}^{\gamma \Rpi^2} = \int \dd \tau \alpha \epsilon \Mp^2 a^2(\tau) \int\frac{\dd^3 \vec{k} \dd^3 \vec{q}}{(2\pi)^6} \underbrace{\epsilon^{ij}_\lambda(\vec{k})  q_i q_j}_{Q_\lambda(\vec{k},\vec{q})} \gamma_\lambda^{\vec{k}}(\tau) \mathcal{R}_{\pi_0}^{\vec{q}}(\tau) \mathcal{R}_{\pi_0}^{-\vec{k}-\vec{q}}(\tau) \,,
\end{equation}
where $\alpha$ is a dimensionless parameter, generally of order one. The Feynman diagram in the left panel of Fig.~\ref{fig: supersolid} is the one corresponding to the contribution under scrutiny. The total tensor power spectrum reads
\begin{align}
    \mathcal{P}_\gamma &= \mathcal{P}_\gamma^\mathrm{vac} + \mathcal{P}_\gamma^{\mathrm{nl}} \,, \,\,\text{with} \\
    \sum_{\lambda,\lambda^\prime}\Braket{\hat{\gamma}^\lambda_{\vec{k}}\hat{\gamma}^{\lambda^\prime}_{\vec{k}^\prime}} &= (2\pi)^3 \delta^{(3)}\left(\vec{k}+\vec{k}^\prime\right) \frac{2\pi^2}{k^3} \mathcal{P}_\gamma(k)\,.
\end{align}
$\mathcal{P}_\gamma^\mathrm{vac}$ is the usual vacuum contribution, whilst the non-linear, i.e. one-loop, scales as:
\begin{equation}
\label{eq: one-loop PS}
        \mathcal{P}_\gamma^{\mathrm{nl}} =  16 \alpha^2 \epsilon^2 \left(\mathcal{P}_{\mathcal{R}_{\pi_0}}\right)^2 \times I_{c_{s2}} \,.
\end{equation}
Before providing more details on the dimensionless quantity $I_{c_{s2}}$ (let us just anticipate that it scales like $1/c_{s2}$), 
we stress two important related features:

\begin{itemize}
    \item  Given the amplification of the scalar power spectrum of $\Rpi$ for small values of the sound speed $c_{s2}$, the $\Rpi$-sourced contribution to the tensor power spectrum can dominate over the vacuum in a large region of parameter space:
    \begin{equation}
        \frac{\mathcal{P}_\gamma^{\mathrm{nl}}}{\mathcal{P}_\gamma^{\mathrm{vac}}} 
        \sim 10 \times \left(\frac{\alpha^2\gamma^2}{\barP/10}\right)
        \times\left(\frac{\epsilon}{0.01}\right) \times\left(\frac{A_s}{10^{-9}}\right) \left(\frac{0.1}{c_{s2}}\right)^{13} > 1 \,,
    \end{equation}
    where we used benchmark values $\mathcal{P}_{\zetan} \sim A_s \sim 10^{-9}\,, \,\, \alpha \sim \gamma \sim 1\,,\,\,\barP \sim 10 $ and $c_{s2} \sim 0.1$ to give the typical order of magnitude of the relative non-linear contribution.
    
    Note moreover that, as detailed also in \cite{Celoria:2020diz}, having a large tensor power spectrum non-linearly sourced by scalars does not necessarily imply a large scalar power spectrum at one-loop. Indeed, the scalars are not enhanced for a very substantial region of the parameter space, thus justifying the use of vacuum fluctuations as the leading contribution for the scalar power spectrum.

    \item If the contribution to the scalar power spectrum of $\Rpi$ is blue-tilted, $n_s^{(\mathrm{en})} > 1$, then the corresponding non-linear contribution to the tensor power spectrum may be dominated by the blue component at small scales:
    \begin{equation}
       \mathcal{P}_\gamma^{\mathrm{nl}}(k) \underset{k \gg k_*}{=}  \mathcal{P}_\gamma^{\mathrm{nl}}(k_*) \left(\frac{k}{k_*}\right)^{n_t^{\mathrm{nl}}} \,, \quad \text{ with }\,
       n_t^{\mathrm{nl}}= 2\left[n_s^{(\mathrm{en})}-1\right] > 0\,.
    \end{equation}
  
    Such tensor power spectrum then can both comply with current observational constraints on $r$ at CMB scales and display a large enhancement at small scales, those probed by LISA, DECIGO/BBO and other next-generation gravitational-wave experiments.
    For example, the corresponding signal may be observable by LISA if the tensor-to-scalar ratio at CMB scales is $r(k_*) = 0.02$ and the tensor tilt is $n_t^\mathrm{nl} =0.3$. 
    Assuming a more conservative value for the tilt of $n_t^\mathrm{nl} =0.1$, 
     the signal can still be observable by the DECIGO/BBO experiment.
\end{itemize}

Using the in-in formalism, the quantity $I_{c_{s2}}$ defined in Eq.~\eqref{eq: one-loop PS} is found to be:
\begin{align}
\label{eq: def Ics2}
    I_{c_{s2}} =& \,  \int_0^\infty \dd v \int_{|1-v|}^{1+v} \dd u \int_0^{\infty^+} \dd x_1 \left[ \int_{x_1}^{\infty^+} \dd x_2 F^\nabla_{c_{s2}}(u,v,x_1,x_2) +
    \int_0^{\infty^-} \dd x_2 F^{\square}_{c_{s2}}(u,v,x_1,x_2) \right] \,, \text{ with } \\
    F^\nabla_{c_{s2}} =& \, - \frac{[4v^2-(1+v^2-u^2)^2]^2}{ 32 u^2v^2}(x_1 x_2)^{-2} \mathrm{Re}\left\{(1-ix_1)(1+ic_{s2} ux_1)(1+ic_{s2} v x_1)
     \right. \nonumber \\
    & \times (1-ic_{s2} ux_2)(1-ic_{s2} vx_2)(1-ix_2)
    \left.\mathrm{exp}[-i x_1(c_{s2}u+c_{s2}v-1)+i x_2(1+c_{s2}u+c_{s2}v)]\right\}  \nonumber \,, \\
        \mathrm{and} \,\, F^\square_{c_{s2}} =& \, \frac{1}{2} \frac{[4v^2-(1+v^2-u^2)^2]^2}{ 32 u^2v^2}(x_1 x_2)^{-2} \left\{(1-ix_1)(1-ic_{s2} ux_1)(1-ic_{s2} v x_1)
        \right. \nonumber \\
    & \times (1+ic_{s2} ux_2)(1+ic_{s2} vx_2)(1+ix_2) 
    \left.\mathrm{exp}[i x_1(c_{s2}u+c_{s2}v+1)-i x_2(1+c_{s2}u+c_{s2}v)]\right\} \nonumber  \,.
\end{align}
Here, $F^\nabla$ corresponds to the integration over the triangle in the time domain, where both the $x_1, x_2$ time contours of integration are deformed in the past in the same direction, while $F^\square$ comes from the integration over the square domain and has a mixed regularisation in the past: $ \pm i \epsilon$ for respectively $x_1, x_2$.
The other two integrals over the variables $(u,v)$ correspond to the remaining integral over the loop momentum $\vec{q}$, whose three components have been decomposed with respect to the external momentum $\vec{k}$ of the tensor mode: $v=q/k$, $u=|\vec{q}-\vec{k}|/k$, the remaining one being the angle $\phi=\mathrm{cos}^{-1}\left[\vec{q}\cdot \vec{k} / (q k) \right]$ over which the integration has already been done explicitly.

The four integrals can be computed analytically thanks to the simple form of the $\Rpi$ modes, which are simple ``massless'' Hankel functions.
Details of the computation can be found in App.~\ref{app: loop power spectrum}, where we show the non-trivial cancellations of the IR divergences between the contributions of $F^\nabla_{c_{s2}}$ and $F^\square_{c_{s2}}$ after integrating over $(x_1,x_2)$.
Then, in order to compute the loop integral, we change the internal momentum variables,
\begin{equation}
\label{eq: t,s variables}
    u=\frac{1+t+s}{2}, \quad v=\frac{1+t-s}{2}\,, 
\end{equation}
after which we find the $c_{s2}$-dependent term to be:

\begin{equation}
\label{eq: Ics2 final}
    I_{c_{s2}}(\mu)=\frac{41 - 118 c_{s2}^2 + 5 c_{s2}^4}{240 c_{s2}} \mathrm{log}(\mu/H)  \,,
\end{equation}
where $\mu$ is the renormalisation scale for this one-loop correction.

The treatment of this logarithmic running has been the subject of interesting discussions (see, e.g., Refs.~\cite{Weinberg:2005vy,Seery:2007we, Adshead:2009cb,Senatore:2009cf, Xue:2011hm, Miao:2012xc,Chen:2016nrs}).
Following \cite{Senatore:2009cf}, we employed here dimensional regularisation  to arrive at the final result, which is both finite and invariant under dilations\footnote{Whilst this work was in progress we corresponded with the authors of \cite{Comelli:2022ikb}, who kindly shared with us their result for the  $c_{s2}$-dependent contribution to the 1-loop power spectrum  in single-field inflation with a non-trivial speed of sound. Our result as to the overall $c_{s2}$-dependence matches theirs exactly. We refer the reader to \cite{Comelli:2022ikb} for a very interesting discussion on non-single-clockness and  the presence of a log-running in the external momentum.}.

In the limit of a small speed of sound, $c_{s2} \ll 1$, such that the non-linear contribution dominates over the vacuum one, the final result for the tensor power spectrum is therefore:
\begin{equation}
    \mathcal{P}_\gamma^{\mu}(k) = \frac{41}{15c_{s2}} \alpha^2\epsilon^2 \left(\mathcal{P}_{\Rpi}\right)^2 
    \times \left(\frac{k}{k_*}\right)^{n_t^{\mathrm{nl}}}
    \times \mathrm{log}\left(\mu/H\right) \,.
\end{equation}

\subsection{Scalar-tensor-tensor bispectrum}

The one-loop STT bispectrum is itself enhanced by the presence of the large $\Rpi$-fluctuations. The fact that this is the case in the squeezed configuration will be particularly relevant for GW anisotropies.

The specific diagram we are considering is represented in the middle panel of Fig.~\ref{fig: supersolid}. It consists of a loop contribution with one external $\zetan$-leg and two external $\gamma$'s. In supersolid inflation, there is a plethora of possible cubic scalar interactions including the $\zetan$ and $\Rpi$ fields. In principle all these are relevant for the computation of the bispectrum. One ought to also remember that the two corresponding quantum operators have non-zero cross-correlation.
Our aim here is to evaluate the bispectrum contribution from one typical interaction term. As we shall see, the parameter space of the model can support a significant STT non-Gaussian signal and a correspondingly large GW anisotropy. We focus on the following cubic term: 
\begin{equation}
    \int \dd t H_\mathrm{int}^{\Rpi^3} = - \int \dd \tau O_1 a(\tau) \epsilon \frac{\Mp^2}{H}   \int\frac{\dd^3 \vec{k} \dd^3 \vec{q}}{(2\pi)^6} \tilde{\mathcal{R}}^{\prime,\vec{k}}_{\pi_0}(\tau) \tilde{\mathcal{R}}^{\prime,\vec{q}}_{\pi_0}(\tau)
    \tilde{\mathcal{R}}^{\prime,-\vec{k}-\vec{q}}_{\pi_0}(\tau)\,, \, \text{ with } \,\,  \tilde{\mathcal{R}}^\prime_{\pi_0} = \mathcal{R}^\prime_{\pi_0} + 3 \mathcal{H} \mathcal{R}_{\pi_0} \,,
\end{equation}
and $O_1$ is  an effective parameter in the cubic scalar Lagrangian whose size is constrained by the current bounds on scalar non-Gaussianity at CMB scales: $O_1 \lesssim  10^{-3} \left( c_{s2} /0.1 \right)^6$~\cite{Celoria:2021cxq}.

In the in-in integrals (see App.~\ref{app: loop mixed bispectrum} for more details), we have to include three vertices, including two nonequivalent cubic interactions, $H_\mathrm{int}^{\Rpi^3}$ and $H_\mathrm{int}^{\gamma^2 \Rpi}$.
For example, one of the six terms reads:
\begin{align}
\label{eq: first term of one-loop bispectrum}
    \Braket{\hat{\gamma}^\lambda_{\vec{k}_1}\hat{\gamma}^{\lambda^\prime}_{\vec{k}_2}\hat{\zetan}_{\vec{k}_3}}_{1\mathrm{-loop}}^{\Rpi^3} = &\int^0_{-\infty^-} \dd \tau_1 a(\tau_1) \int^{\tau_1}_{-\infty^-} \dd \tau_2 a(\tau_2)
    \int^{\tau_2}_{\infty^-} \dd \tau_3 a(\tau_3) \nonumber \\
    & \times \mathrm{Im} \left\{
    \Braket{0|
    \hat{H}_\mathrm{int}^{\Rpi^3}(\tau_3)
    \hat{H}_\mathrm{int}^{\gamma^2\Rpi}(\tau_2)
    \hat{H}_\mathrm{int}^{\gamma^2\Rpi}(\tau_1)
    \hat{\gamma}^\lambda_{\vec{k}_1}\hat{\gamma}^{\lambda^\prime}_{\vec{k}_2}\hat{\zetan}_{\vec{k}_3}
    |0} \right\}\,.
\end{align}
In the limit of interest for us, that of a soft scalar mode, we are able to perform the calculation analytically and find:
\begin{align}
\label{eq: one-loop BS}
    \sum_{\lambda,\lambda^\prime}\Braket{\hat{\gamma}^\lambda_{\vec{k}_1}\hat{\gamma}^{\lambda^\prime}_{\vec{k}_2}\hat{\zetan}_{\vec{k}_3}}_{1\mathrm{-loop}}^{\Rpi^3} = (2\pi)^3 &\delta^{(3)}\left(\vec{k}_1+\vec{k}_2+\vec{k}_3\right) B^{\gamma\gamma\zeta_n}_{1\mathrm{-loop}}(k_1,k_2,k_3)\,, \quad\text{ with } \nonumber \\
    B^{\gamma\gamma\zeta_n}_{1\mathrm{-loop}}(k_S,k_S,k_L) \underset{k_L \ll k_S}{=}
    - 24 \pi^2 O_1 \epsilon &\frac{\Mp^2}{H^2} \mathcal{P}_{\Rpi}
    \times P_{\Rpi\zetan} (k_L) \times \frac{P_\gamma^{\mathrm{nl}}(k_S)}{I_{c_{s2}}} \times J_{c_{s2}} \,.
\end{align}
Before turning to the details on $J_{c_{s2}}$,  
let us estimate the size of the related $f_\mathrm{NL}$ parameter:

\begin{equation}
    f_{\mathrm{NL},\mathrm{sq}}^{\gamma\gamma\zetan} = \frac{B^{\gamma\gamma\zeta_n}_{1\mathrm{-loop}}(k_S,k_S,k_L \ll k_S)}{P_\gamma(k_S) P_{\zetan} (k_L)} \sim O_1 \frac{\mathcal{P}_{\Rpi}}{\mathcal{P}_0} \frac{\mathcal{P}_{\zetan\Rpi}}{\mathcal{P}_{\zetan}}
    \frac{J_{c_{s2}}}{I_{c_{s2}}} \,,
\end{equation}
where we focused on the regime where the tensor power spectrum is dominated by the non-linear contribution: $P_\gamma \simeq P_\gamma^{\mathrm{nl}}$.
We expect the first two ratios to be typically much larger than one due to the relative enhancement of the $\Rpi$-fluctuations, which can easily counter-balance the smallness of $O_1$.
It is interesting to notice that there is enough freedom in supersolid inflation to accommodate  relatively small scalar non-Gaussianities (i.e. compatible with observational constraints) and, at the same time, a large STT bispectrum.

The explicit expression for the dimensionless quantity $J_{c{s2}}$, defined in Eq.~\eqref{eq: one-loop BS}, is
\begin{align}
\label{eq: def Jcs2}
    J_{c_{s2}} &= \int_0^\infty \dd v \int_{|1-v|}^{1+v} \dd u \int_0^{\infty^+} \dd x_1 \int_{x_1}^{\infty^+} \dd x_2
    \left[ \int_{x_2}^{\infty^+} \dd x_3   \, G_{c_{s2}}^\nabla(x_i,u,v)  + \int_{0}^{\infty^-} \dd x_3   \, G_{c_{s2}}^\square(x_i,u,v)  \right]\,,
\end{align} 
where the two contributions $G_{c_{s2}}^{\nabla / \square}$ correspond to different time orderings of the interaction Hamiltonians with respect to the external operators.
Each of the two contributions contains three independent terms.
For example, the first term in $G^\nabla$, corresponding to Eq.~\eqref{eq: first term of one-loop bispectrum}, reads:
\begin{align}
\label{eq: first term of Gcs2}
    G^{\nabla, 1}_{c_{s2}} &= - 3  \frac{\left[4 v^2 -(1+v^2+u^2)^2\right]^2}{32u^2v^5} x_3^{-4} x_2^{-2} x_1^{-2} \mathrm{Im} \left\{(1-ix_1)(1+ivc_{s2}x_1)(1+iuc_{s2}x_1)
    (1-ix_2)
    \right.  \\
    & \left. (1+ivc_{s2}x_2)(1-iuc_{s2}x_2)  \left[v^2 c_{s2} x_3^2 - 3\left(1-ivc_{s2}x_3\right)\right]^2 e^{i[1-(v+u)c_{s2}]x_1
   +i[1-(v-u)c_{s2}]x_2 
   +i 2v c_{s2}x_3}\right\} \nonumber  \,.
\end{align}
When calculating the triple time integrals (over $x_i$) for each of the 6 contributions, one finds that although individual terms carry up to $5$ different kinds of IR divergences in the limit $x_i \rightarrow 0$,  they all cancel once summed together (see App.~\ref{app: loop mixed bispectrum} for  more details). One then moves to the  integrals over $(u,v)\rightarrow(t,s)$ to find, after dimensional regularisation, the following expression for $J_{c_{s2}}$,
\begin{align}
\label{eq: Jcs2 final}
    J_{c_{s2}}(\mu) &= \frac{553455+174300 c_{s2}^2 - 328 062 c_{s2}^4 - 1 325 076 c_{s2}^6 + 65 095 c_{s2}^8}{430080 c_{s2}^2}\mathrm{log}(\mu/H)  \nonumber \\
    &\underset{c_{s2}\rightarrow 0} {\longrightarrow} \frac{5271}{4096  c_{s2}^2} \mathrm{log}(\mu/H)\,.
\end{align}
Note that the running of the one-loop mixed bispectrum is exactly the same than the one of the one-loop power spectrum.
It follows that the related $f_\mathrm{NL}$ parameter has a simple expression independent of $\mu$, in the limit $c_{s2} \rightarrow 0$ it reads:
\begin{align}
    f_{\mathrm{NL},\mathrm{sq}}^{\gamma\gamma\zetan}  &= - 3 O_1 
    \frac{\beta/\bar{\mathcal{P}}}{c_{s2}^3}
    \frac{\gamma}{c_{s2}^6}
    \frac{79065}{10496c_{s2}}\,.
\end{align}

The constraints from the scalar sector  notwithstanding (i.e. $O_1 \lesssim 10^{-3} \left( c_{s2} /0.1 \right)^6$), the parameter space of the model leaves ample room for a  large STT squeezed bispectrum:
\begin{align}
\label{eq: fNL final}
    f_{\mathrm{NL},\mathrm{sq}}^{\gamma\gamma\zetan}  \sim
    10^6 \times \left(\frac{O_1}{10^{-3}(c_{s2}/0.1)^6}\right) \times \left(\frac{\beta \gamma}{\bar{\mathcal{P}}/10}\right) \times \left( \frac{0.1}{c_{s2}}\right)^4 \,.
\end{align}
Note that the numerical value $10^6$ is arrived at by  saturating the observational constraint on scalar non-Gaussianities and taking $(\beta \gamma /\bar{\mathcal{P}}) \sim 0.1$, $c_{s2} \sim 0.1$. This goes to show that one may well have $f_{\mathrm{NL},\mathrm{sq}}^{\gamma\gamma\zetan} \gg 1$ in the model at hand while complying with observational constraints in the scalar sector.

\subsection{Primordial gravitational wave anisotropies}

As we have seen in the previous cases, here too the existence of a STT squeezed bispectrum induces an anisotropic component on the GW spectrum.

Indeed, the effect of a long scalar mode is well-described by the action of a classical background source on the short tensor two-point function.
We see this at the level of the in-in calculation where the long scalar mode $\Rpi$ during inflation becomes
\begin{equation}
    \hat{\tilde{\mathcal{R}}}^\prime_{\pi_0} \longrightarrow \tilde{\mathcal{R}}_{\pi_0}^{\mathrm{cl},\prime} \simeq  3 \mathcal{H}\Rpi^{\mathrm{cl}} \,.
\end{equation}
Note that  we have consistently neglected the terms suppressed on (super-) horizon scales.
A typical contribution (there are six in total)  reads
\begin{align}
   \label{44a}
    \left.\Braket{\hat{\gamma}^\lambda_{\vec{k}_1}\hat{\gamma}^{\lambda^\prime}_{\vec{k}_2}}\right|_{\Rpi^\mathrm{cl}} = & \, 
    \int^0_{-\infty^-} \dd \tau_1 a(\tau_1) \int^{\tau_1}_{-\infty^-} \dd \tau_2 a(\tau_2)
    \int^{\tau_2}_{\infty^-} \dd \tau_3 a(\tau_3) \nonumber \\
    & \times \mathrm{Im} \left\{
    \Braket{0|\hat{H}_\mathrm{int}^{\Rpi^2 \Rpi^\mathrm{cl}}(\tau_3)
    \hat{H}_\mathrm{int}^{\gamma^2\Rpi}(\tau_2)
    \hat{H}_\mathrm{int}^{\gamma^2\Rpi}(\tau_1)
    \hat{\gamma}^\lambda_{\vec{k}_1}\hat{\gamma}^{\lambda^\prime}_{\vec{k}_2}|0} \right\}  \,,
\end{align}
where, in the momentum configuration $|\vec{k}_1+\vec{k}_2|\ll k_1,k_2$,
the real classical source may be factored out 
of the expectation value, the imaginary part, and the integral sign.

\begin{figure}
    \centering
     \includegraphics[scale=0.5]{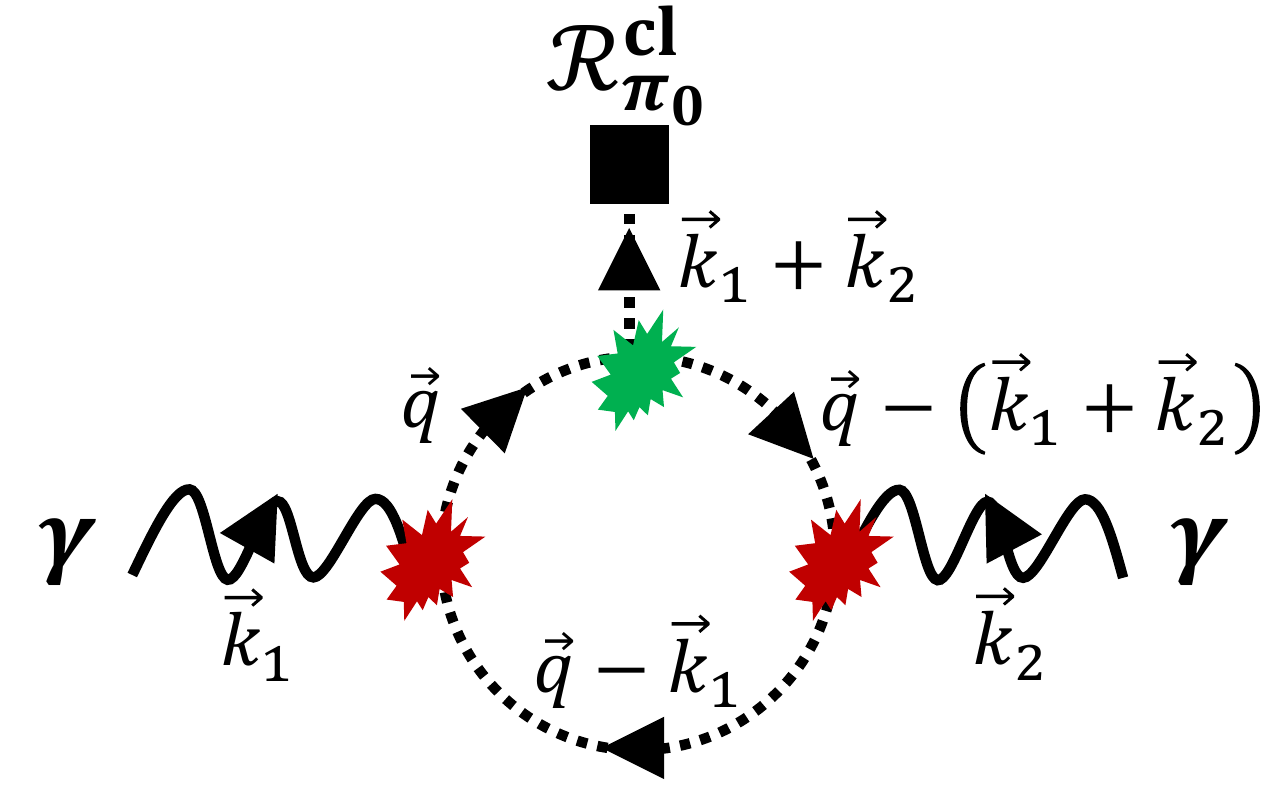}
    \caption{Momenta flowing in the sourced tensor two-point function.}
    \label{fig: supersolid momenta}
\end{figure}

Let us provide some more details. The integral in $\dd \tau_3$, the one associated to the purely scalar vertex, counts three functions whose arguments are (see Fig.~\ref{fig: supersolid momenta}) $q\tau_3$ and $|\vec{q}-(\vec{k}_1+\vec{k}_2)|\tau_3$ inside the loop, and $|\vec{k}_1+\vec{k}_2|\tau_3$ outside it.
Given the oscillating behaviour of all these functions inside the horizon, the effective\footnote{That is, the domain of integration over which one does not have those fast oscillations that lead to a strong suppression.} domain of integration is set by the horizon of the function with the largest wavenumber.
It is immediate to see that, upon implementing the $|\vec{k}_1+\vec{k}_2|\ll k_1,k_2$ hierarchy, the function $\mathcal{R}_{\pi_0,-\vec{k}_1-\vec{k}_2}^\mathrm{cl}$ can, with one exception, be moved outside the time-integral as it is well-approximated by a constant throughout the entire effective domain of integration.
The exception is for those values of the variable $\vec{q}$ such that $\vec{q} \simeq \vec{0}$ or $\vec{q}\simeq \vec{k}_1+\vec{k}_2$.
This represents an extremely small domain over the momentum loop integral given that it corresponds to a precise value for both the norm and the orientation of the vector $\vec{q}$. 

Furthermore, we know that this very same momentum configuration will receive negligible contribution from the two remaining tensor-scalar-scalar vertices of the 1-loop calculation of Eq.(\ref{44a}).
This is immediate to see upon recalling that  both the interaction and the wavefunctions in the TSS vertices are the same ones as in single field slow-roll.
Inspecting the momentum configuration that contribute to the TSS calculation in SFSR (see e.g. Fig.~2 of \cite{Meerburg:2016ecv}) makes it clear that the configuration corresponding to  having a very soft scalar in the loop can be safely disregarded. The function $\mathcal{R}_{\pi_0}^\mathrm{cl}$ then can be factored out of the integral and treated as a background.

\bigskip

Comparing the resulting expression with Eq.~\eqref{eq: first term of one-loop bispectrum}, we find:
\begin{align}
    \sum_{\lambda,\lambda^\prime}\left.\Braket{\hat{\gamma}^\lambda_{\vec{k}_1}\hat{\gamma}^{\lambda^\prime}_{\vec{k}_2}}\right|_{\Rpi^\mathrm{cl}} \underset{|\vec{k}_1 + \vec{k}_2| \ll k_1,k_2}{=} & \frac{\mathcal{R}^{\mathrm{cl}}_{\pi_0,\vec{k}_1+\vec{k}_2}}{P_{\Rpi \zetan}(|\vec{k}_1 + \vec{k}_2|)} B^{\gamma\gamma\zeta_n}_{1\mathrm{-loop}}(k_1,k_2,|\vec{k}_1 + \vec{k}_2|) \,.
\end{align}
In order to compare this result with previous literature, we can rewrite this expression for the GW anisotropy in supersolid inflation as
\begin{equation}
\label{eq: final anisotropies}
    \sum_{\lambda,\lambda^\prime}\left.\Braket{\hat{\gamma}^\lambda_{\vec{k}_1}\hat{\gamma}^{\lambda^\prime}_{\vec{k}_2}}\right|_{\Rpi^\mathrm{cl}} \underset{|\vec{k}_1 + \vec{k}_2| \ll k_1,k_2}{=} \int%_{q \ll k_1,k_2}
    \dd^3 \vec{q} \, \delta^{(3)} (\vec{q}+\vec{k}_1+\vec{k}_2) \frac{B^{\gamma\gamma\zeta_n}_{1\mathrm{-loop}}(k_1,k_2,q) }{P_{\Rpi \zetan}(q)} \mathcal{R}^{\mathrm{cl}}_{\pi_0,-\vec{q}} \,.
\end{equation}
This is the STT analogous of the TTT result found in \cite{Dimastrogiovanni:2019bfl}. 
The latter was written down for an unspecified, phenomenological, model and arrived at in a heuristic fashion.
Here  we provided a clear path to the result via the in-in formalism by (i) calculating the 1-loop STT bispectrum in supersolid inflation and (ii) showing its relation with the induced anisotropy on the GW power spectrum.

Note that in the expression of Eq.~\eqref{eq: final anisotropies}, both the LHS and RHS are evaluated at the end of inflation.
Therefore, even though $\Rpi$ may not survive the reheating process -- we remind that it is $\zetan$ that is transmitted to $\zeta$ that then seeds the temperature anisotropies of the CMB -- the super-horizon tensor modes are frozen until they re-enter the horizon (for example, modes relevant for LISA re-enter during radiation-domination), and still carry the information about the value of the $\left(\mathcal{R}^{\mathrm{cl}}_{\pi_0,-\vec{q}}\right)$-fluctuations during inflation.
Then, the tensor power spectrum evolves on small scales as usual, but with different values in different patches of the sky due to different realisations of the stochastic variable $\Rpi^\mathrm{cl}$, therefore giving rise to SGWB anisotropies of primordial origin, $\delta_\mathrm{GW}(k,\hat{n})$.

These anisotropies of primordial origin in the SGWB have a typical amplitude
\begin{align}
    \sqrt{\braket{\delta_\mathrm{GW}^2(k,\hat{n})} } &\sim   f_{\mathrm{NL},\mathrm{sq}}^{\gamma\gamma\zetan}\left(\frac{\sqrt{\mathcal{P}_{\zetan}\mathcal{P}_{\Rpi}}}{\mathcal{P}_{\zetan\Rpi}}\right)_{k_*} A_s^{1/2} \,,
\end{align}
where $k_*$ is the CMB pivot scale.
Given the ample room for a large mixed three-point function in the squeezed limit, even taking into account observational constraints in the scalar sector, see Eq.~\eqref{eq: fNL final}, it is certainly possible to have percent-level anisotropies in supersolid inflation.

\bigskip

We also note in passing that, in supersolid inflation, GW anisotropies $\delta_{\text{GW}}$ have a non-zero cross-correlation (see \cite{Adshead:2020bji}, \cite{Dimastrogiovanni:2021mfs} for the first work and related applications on this topic respectively) with CMB temperature anisotropies, because in such model $\Rpi$ (sourcing $\delta_{GW}$) and $\zetan$ (sourcing $\delta T$) are themselves correlated:
\begin{equation}
    \Braket{\delta_{\text{GW}}(k,\hat{n}) \delta T(k_*,\hat{n})} \propto \Braket{\mathcal{R}^{\mathrm{cl}}_{\pi_0,-\vec{k}_*} \zeta^\mathrm{cl}_{n,\vec{k}_*}} \propto \mathcal{P}_{\Rpi\zetan}(k_*) \,. %\text{\textcolor{red}{ $ \gg A_s$}} \,.
\end{equation}
We do not provide here further details on cross-correlations in this model, leaving the subject for future work.
Having tackled the one-loop calculation of the tensor power spectrum and STT bispectrum, one might wonder about higher order corrections, starting with the two-loop term. As attested by the 1-loop result, the two-loop calculation is bound to be rather complex and, for this reason, we shall leave it for future work. Nevertheless, one may  note that in several setups the two-loop contribution can be (parametrically or by other mechanisms) sub-leading w.r.t. to the one loop term whilst having, at the same time, that the one loop correction is larger than the tree level contribution. Two independent interesting examples of this sort in the inflationary context can be found in \cite{Pearce:2017bdc,Barnaby:2012tk}.

\section{Conclusions}
\label{conclusions}

The prospect of a near-future detection of the stochastic gravitational wave background holds an immense discovery potential for the physics of the early universe. Already the spectral shape and the chirality of the GW signal may reveal key information on, for example, the inflationary particle content. We can probe even deeper into the physics of the early-time acceleration as we test primordial non-Gaussianities, directly accessing inflationary interactions.
A remarkable handle on such non-Gaussianities consists in the anisotropy these induce on the GW power spectrum. Indeed, the squeezed component of a non-trivial scalar(tensor)-tensor-tensor bispectrum  can be constrained by surveying the position dependence of the tensor power spectrum \cite{Jeong:2012df,Dai:2013kra,Dimastrogiovanni:2019bfl}. The same reasoning applies to non-Gaussianity-induced anisotropies of the scalar power spectrum, with intriguing applications \cite{Jeong:2012df,Dai:2013kra} to the physics of the CMB and the large scale structure.

In this work, we set out to put these results on a firmer ground, deriving them by means of the in-in formalism. We first introduced the framework and then made contact with  existing examples. Models whose GW signal may exceed the sensitivity threshold of upcoming experiments include the EFT of non-minimally coupled spinning (spin-2 in particular) fields \cite{Bordin:2017ozj} and supersolid inflation \cite{Celoria:2020diz}. We have gone beyond the existing work on the spin-2 case by showing formally how to derive the STT bispectrum and related GW anisotropy beyond the small-mixing regime. 
In order to do the same for the supersolid setup, we calculated for the first time the 1-loop STT bispectrum. Our treatment clarifies under what conditions specific inflationary interactions give rise to GW anisotropies. 

The inflationary models of interest are those, typically multi-field, able to support a detectable GW spectrum and to produce a sufficiently large long-short mode coupling to induce a sizable anisotropy. As an efficient ``rule of thumb'' when casting inflationary mechanisms for the part, it is useful to consider those characterised by light fields with non-derivative interactions. 

It is important to note that there exist anisotropies of a different nature that may nevertheless exhibit the same signature as those studied in this manuscript. The propagation of standard tensor modes through structure, in their way to GW detectors, will itself result in an anisotropic component \cite{Contaldi:2016koz,Bartolo:2019oiq}, which is comparable to those originating from non-Gaussianities when the strength of the latter is, schematically, $f_{\rm NL}\sim 1$. The models we studied here can support a much larger $f_{\rm NL}$, but one ought to be aware of the possible degeneracies for relatively small non-Gaussianities. There is of course also the matter of anisotropies of astrophysical origin, a subject which is currently under intense research scrutiny \cite{Cusin:2017fwz,Cusin:2018rsq,Jenkins:2018kxc,Cusin:2019jhg,Bertacca:2019fnt,Pitrou:2019rjz,Bellomo:2021mer}.

In this context, an intriguing possibility to single out  the cosmological nature of anisotropies relies on cross-correlation with the CMB \cite{Adshead:2009cb,Ricciardone:2021kel}, an option available both for anisotropies associated to STT and TTT primordial bispectra. Having focused mostly on  bispectra of the STT type in this work, a natural next step would be to consider such cross-correlation, for example for the supersolid case, and forecast the  bounds on squeezed non-Gaussianities one may achieve with a specific GW experiment. We leave this to future work.

\subsection*{Acknowledgements}
We are indebted to Denis Comelli, Maicol Di Giambattista, Luigi Pilo, and Rocco Rollo for insightful discussions and comments as well as for sharing a draft of their work \cite{Comelli:2022ikb} before publication.
We would also to like to thank Sebastian Garcia-Saenz for interesting discussions about the use of the renormalization scale in cosmology.
We acknowledge support from the International Emerging Action between Institut d'Astrophysique de Paris (SU-CNRS) and the University of Groeningen.
M.F. and L.P. would like to acknowledge support from the “Atracci\'{o}n de Talento” grant 2019-T1/TIC15784, their work is partially supported by the Spanish Research Agency (Agencia Estatal de Investigaci\'{o}n) through the Grant IFT Centro de Excelencia Severo Ochoa No CEX2020-001007-S, funded by MCIN/AEI/10.13039/501100011033.

\appendix

\section{Standard in-in formalism}
\label{ininstandard}

Given a theory, defined in terms of a Lagrangian for a set of fields $\psi_i$, one divides the corresponding Hamiltonian into a free part and interactions: $H=H_\mathrm{free}+H_\mathrm{int}$.
The free part defines what we call the interaction picture fields $\psi_i^I$ during inflation, and that verify the equations of motion derived from $H_\mathrm{free}$:
\begin{equation}
    \left.\frac{\delta H_\mathrm{free}}{\delta \psi_i}\right|_{\psi_i^I} = 0 \,.
\end{equation}
Then, interactions encoded in $H_\mathrm{int}$ are treated perturbatively, and one computes correlation functions of the full fields $\psi_i$ in an expansion in $H_\mathrm{int}$.
The master formula of the in-in formalism enables to express the time-dependent vacuum expectation value of a quantum operator $\hat{\mathcal{O}}(\hat{\psi}_i)$ made of the fundamental fields, in the following way:
\begin{equation}
\label{eq: in-in master formula}
\boxed{
    \Braket{\hat{\mathcal{O}}(t)} = \Braket{0|\bar{T} \left(e^{i \int_{-\infty^+}^t \dd t^\prime \hat{H}_\mathrm{int}^I(t^\prime)} \right) \hat{\mathcal{O}}^I(t) T \left(e^{-i \int_{-\infty^-}^t \dd t^{\prime\prime} \hat{H}_\mathrm{int}^I(t^{\prime\prime})}   \right) |0 } \,,
}
\end{equation}
where $\hat{H}_\mathrm{int}^I=\hat{H}_\mathrm{int}(\hat{\psi}_i^I)$ and $\hat{\mathcal{O}}^I=\hat{\mathcal{O}}(\hat{\psi}_i^I)$.
The symbols $-\infty^\pm$ represent the fact that the adiabatic vacuum of the interacting theory, $\ket{0}$, should be projected onto the $\ket{\mathrm{in}}$ vacuum of the free theory in the infinite past.
This procedure can be implemented technically by slightly deforming the contour of integration into the complex plane in the infinite past, $-\infty \rightarrow -\infty(1\pm i \epsilon)$, hence the notations.
The symbol $T$ (and its hermitian conjugate $\bar{T}$), represents the (anti-)time ordered operator.
Note that the $i \epsilon$-prescriptions are conjugate one to the other in the two branches of the closed contour of integration, and that this is required to preserve the hermiticity of the full evolution operator from the free vacuum to an arbitrary interacting state at time $t$,
\begin{equation}
    \hat{U} (t)  = \bar{T} \left( e^{i \int_{-\infty^+}^t \dd t^\prime \hat{H}_\mathrm{int}^I(t^\prime)} \right) \,.
\end{equation}
This technical detail will turn out to be important for concrete calculations using the in-in formalism, but in more formal steps in the following we may drop this explicit boundary dependence.

In practice, correlation functions like $\Braket{\hat{\mathcal{O}}(t)}$ are computed perturbatively, at a given order in $H_\mathrm{int}$, which can each be represented by a Feynman diagram for cosmological correlators.
The first terms of the expansion can easily be written explicitly.
Defining $\Braket{\hat{\mathcal{O}}(t)}^{(n)}$ the $n$-th order of the expansion, one has:
\begin{align}
    \Braket{\hat{\mathcal{O}}(t)}^{(0)} =& \Braket{0| \hat{\mathcal{O}}^I(t) |0}  \\\label{formula1}
    \Braket{\hat{\mathcal{O}}(t)}^{(1)} =& - 2 \mathrm{Im}\left[\int_{-\infty^+}^t \dd t^\prime \Braket{0| \hat{H}^I_\mathrm{int}(t^\prime) \hat{\mathcal{O}}^I(t)  |0} \right]\\
    \Braket{\hat{\mathcal{O}}(t)}^{(2)} =& - 2 \mathrm{Re}\left[ \int_{-\infty^+}^t \dd t^\prime \int_{-\infty^+}^{t^\prime} \dd t^{\prime\prime} \Braket{0|\hat{H}^I_\mathrm{int}(t^{\prime\prime}) \hat{H}^I_\mathrm{int}(t^\prime)\hat{\mathcal{O}}^I(t)|0} \right] \\
    & + \int_{-\infty^+}^t \dd t^\prime  \int_{-\infty^-}^t \dd t^{\prime\prime} \Braket{0| \hat{H}^I_\mathrm{int}(t^{\prime}) \hat{\mathcal{O}}^I(t)  \hat{H}^I_\mathrm{int}(t^{\prime\prime})|0} \nonumber \\
    \cdots & \nonumber \\
    \Braket{\hat{\mathcal{O}}(t)}^{(n)} =&\, i^n \int_{-\infty}^t \dd t_1 \int_{-\infty}^{t_1} \dd t_2 \ldots \int_{-\infty}^{t_{n-1}} \dd t_n \\
    & \times \Braket{0|\left[\hat{H}_\mathrm{int}^I(t_n), \left[ \hat{H}_\mathrm{int}^I(t_{n-1}),  \ldots, \left[ \hat{H}_\mathrm{int}^I(t_1) , \hat{\mathcal{O}}^I(t)\right]  \ldots \right]  \right]|0} \nonumber \,.
\end{align}
The way one wrote the $n$-th order term is called the nested commutator form, it can be obtained by mixing terms like the first and second lines of $\Braket{\hat{\mathcal{O}}(t)}^{(2)}$.
There, identifying $-\infty^+$ with $-\infty^-$, one can express the integral over a square in the second line (indeed, $t^\prime$ and $t^{\prime\prime}$ are then interchangeable), as two integrals over a triangle in the time domain:
\begin{align*}
    &\int_{-\infty}^t \dd t^\prime  \int_{-\infty}^t \dd t^{\prime\prime} \Braket{0| \hat{H}^I_\mathrm{int}(t^{\prime}) \hat{\mathcal{O}}^I(t)  \hat{H}^I_\mathrm{int}(t^{\prime\prime})|0} \\
    &= \int_{-\infty}^t \dd t^\prime \int_{-\infty}^{t^\prime} \dd t^{\prime\prime} \Braket{ 0| \left(\hat{H}^I_\mathrm{int}(t^{\prime}) \hat{\mathcal{O}}^I(t)  \hat{H}^I_\mathrm{int}(t^{\prime\prime})+ \hat{H}^I_\mathrm{int}(t^{\prime\prime}) \hat{\mathcal{O}}^I(t)  \hat{H}^I_\mathrm{int}(t^{\prime}) \right) |0} \,,
\end{align*}
and therefore:
\begin{equation}
    \Braket{\hat{\mathcal{O}}(t)}^{(2)} = - \int_{-\infty}^t \dd t^\prime \int_{-\infty}^{t^\prime} \dd t^{\prime\prime} \Braket{0|\left[ \hat{H}^I_\mathrm{int}(t^{\prime\prime}) , \left[ \hat{H}^I_\mathrm{int}(t^\prime) , \hat{\mathcal{O}}^I(t) \right] \right]|0} \,.
\end{equation}
Note that the price for this compact rewriting is loosing the information about the $i \epsilon$-prescriptions that shut off interactions in the infinite past.
Therefore, the nested commutator formula is more useful for formal derivations than practical computations, for which one would encounter UV divergences difficult to regularize without coming back to the un-nested form.

\section{One-loop calculations in supersolid inflation}
\label{app: loop calculations}

In this appendix, we give details about the one-loop calculations of the tensor power spectrum and STT bispectrum in supersolid inflation.

\subsection{One-loop tensor power spectrum}
\label{app: loop power spectrum}

Using the in-in formula and Eq.~\eqref{eq: TSS Hamiltonian supersolid}, we can express the tensor two-point function as:
\begin{align}
    \Braket{\hat{\gamma}^\lambda_{\vec{k}}\hat{\gamma}^{\lambda^\prime}_{\vec{k}^\prime}} = & \epsilon^2 \alpha^2 \Mp^4
    \int \frac{\dd^3 \vec{k}_1\dd^3 \vec{q}_1\dd^3 \vec{k}_2\dd^3 \vec{q}_2}{(2\pi)^{12}}
    \sum_{\bar{\lambda},\bar{\lambda}^\prime}
    Q_{\bar{\lambda}}(\vec{k}_1,\vec{q}_1)
    Q_{\bar{\lambda}^\prime}(\vec{k}_2,\vec{q}_2)
    %\epsilon^{ij}_{\bar{\lambda}}(\hat{k}_1)q_1^iq_1^j
    %\epsilon^{kl}_{\bar{\lambda}^\prime}(\hat{k}_2)q_2^kq_2^l
    \\
    & 
    \Biggl[ 
    - 2 \mathrm{Re} 
    \Biggl(     \int_{-\infty^+}^0 \dd \tau_1 a^2(\tau_1)
    \int_{-\infty^+}^{\tau_1} \dd \tau_2 a^2(\tau_2) 
    \nonumber\\
    & \quad\quad\quad\quad
    \Braket{0|
    \hat{\gamma}^{\bar{\lambda}}_{\vec{k}_2}(\tau_2)
    \hat{\mathcal{R}}_{\pi_0}^{\vec{q}_2}(\tau_2)
    \hat{\mathcal{R}}_{\pi_0}^{-\vec{k}_2-\vec{q}_2}(\tau_2)
    \hat{\gamma}^{\bar{\lambda}^\prime}_{\vec{k}_1}(\tau_1)
    \hat{\mathcal{R}}_{\pi_0}^{\vec{q}_1}(\tau_1)
    \hat{\mathcal{R}}_{\pi_0}^{-\vec{k}_1-\vec{q}_1}(\tau_1)
    \hat{\gamma}^\lambda_{\vec{k}}
    \hat{\gamma}^{\lambda^\prime}_{\vec{k}^\prime}
    |0}
    \Biggr)\nonumber \\
    & + 
    \int_{-\infty^+}^0 \dd \tau_1 a^2(\tau_1)
    \int_{-\infty^+}^{0} \dd \tau_2 a^2(\tau_2)  \nonumber
    \\
    & \quad\quad\quad\quad 
    \Braket{0|
    \hat{\gamma}^{\bar{\lambda}^\prime}_{\vec{k}_1}(\tau_1)
    \hat{\mathcal{R}}_{\pi_0}^{\vec{q}_1}(\tau_1)
    \hat{\mathcal{R}}_{\pi_0}^{-\vec{k}_1-\vec{q}_1}(\tau_1)
    \hat{\gamma}^\lambda_{\vec{k}}
    \hat{\gamma}^{\lambda^\prime}_{\vec{k}^\prime}
    \hat{\gamma}^{\bar{\lambda}}_{\vec{k}_2}(\tau_2)
    \hat{\mathcal{R}}_{\pi_0}^{\vec{q}_2}(\tau_2)
    \hat{\mathcal{R}}_{\pi_0}^{-\vec{k}_2-\vec{q}_2}(\tau_2)
    |0}
    \Biggr] \nonumber \,,
\end{align}
where $Q_{\lambda}(\vec{k},\vec{q})=\epsilon^{ij}_{\lambda}(\hat{k})q^iq^j$.
Performing all possible Wick contractions, and changing to the dimensionless time variables
\begin{equation}
    x_i = -k \tau_i \longrightarrow a(\tau_i) = \frac{k}{H x_i} \,,
\end{equation}
one arrives at the dimensionless one-loop tensor power spectrum:
\begin{equation}
    \mathcal{P}_\gamma^\mathrm{nl}  = 16 \alpha^2 \epsilon^2 \left(\mathcal{P}_{\Rpi}\right)^2 I_{c_{s2}}\,.
    \end{equation}
This equation is formally the same as the final one in the body of the paper, see Eq.~\eqref{eq: one-loop PS}, but $I_{c_{s2}}$ is not yet in its simplest form,
\begin{align}
\label{eq: Ics2 before simplify}
    I_{c_{s2}} &= \frac{1}{2\pi} \sum_{\lambda,\lambda^\prime}
    \int  \dd^3 \left(\vec{q}/k\right)
    \frac{Q_\lambda(\vec{k},\vec{q})}{k^2}
    \frac{Q_{\lambda^\prime}(\vec{k},\vec{q})}{k^2}
    \frac{k^6}{q^3|\vec{q}-\vec{k}|^3}  \\
    & \quad \times
    \int_0^{\infty^+} \dd x_1
    \left[
    \int_{x_1}^{\infty^+} \dd x_2 f^\nabla_{c_{s2}}(x_1,x_2) +
    \int_0^{\infty^-} \dd x_2 f^{\square}_{c_{s2}}(x_1,x_2) \right] \,, \nonumber
\end{align}
with:
\begin{align}
    f^\nabla_{c_{s2}} =& \, -
    (x_1 x_2)^{-2} \mathrm{Re}\left\{(1-ix_1)(1+ic_{s2} ux_1)(1+ic_{s2} v x_1)
     \right.  \\
    & \times (1-ic_{s2} ux_2)(1-ic_{s2} vx_2)(1-ix_2)
    \left.\mathrm{exp}[-i x_1(c_{s2}u+c_{s2}v-1)+i x_2(1+c_{s2}u+c_{s2}v)]\right\}  \nonumber \,, \\
    f^\square_{c_{s2}} =& \,
    \frac{1}{2}(x_1 x_2)^{-2} \left\{(1-ix_1)(1-ic_{s2} ux_1)(1-ic_{s2} v x_1)
        \right. \nonumber \\
    & \times (1+ic_{s2} ux_2)(1+ic_{s2} vx_2)(1+ix_2) 
    \left.\mathrm{exp}[i x_1(c_{s2}u+c_{s2}v+1)-i x_2(1+c_{s2}u+c_{s2}v)]\right\} \nonumber  \,.
\end{align}
In these expressions, we have already introduced the variables $(u,v)$ for the internal momentum $\vec{q}$ in the loop:
\begin{align}
    v&=\frac{q}{k} \,, \quad 
    u=\frac{|\vec{q}-\vec{k}|}{k} \,, \quad  \text{and}  \quad
    \phi = \mathrm{cos}^{-1}\left(\frac{\vec{q}\cdot \vec{k}}{q k} \right) \,, \quad \text{such that:} \\
    \int \dd^3 \left(\vec{q}/k\right) &\longrightarrow \int_0^\infty \dd v \int_{|v-1|}^{v+1} \dd u \int_{0}^{2\pi} \dd \phi \times uv  \,, \nonumber\\
    \frac{Q_{\lambda}(\vec{k},\vec{q})}{k^2}&\longrightarrow \frac{1}{4\sqrt{2}}\left[4v^2-(1+v^2-u^2)^2\right] \alpha_\lambda(\phi) \nonumber \,.
\end{align}
The properties of the functions $\alpha_\lambda(\phi)$ can be inferred from those of the polarisation tensors. Here we only need the following one:
\begin{equation}
    \int_{0}^{2\pi} \dd \phi  \, \alpha_\lambda(\phi)  \alpha_{\lambda^\prime}(\phi) = \pi \, \delta^{\lambda\lambda^\prime} \,.
\end{equation}
By applying these simplifications in the first line of $I_{c_{s2}}$ in Eq.~\eqref{eq: Ics2 before simplify}, we recover the expression in the main part of this text, see Eq.~\eqref{eq: def Ics2}.

\paragraph{Computation of $I_{c_{s2}}$.}
As we will see, both contributions from the time-integrals of $F^\nabla$ and $F^\square$ are divergent in the IR, i.e. for small values of the time parameter in the integral.
Crucially, these divergences cancel each other such that a finite result may be extracted. We compute first the time-integrals, separating the two contributions and adding a small IR cutoff $x_0$:
\begin{align}
    &\int_{x_0}^{\infty^+} \dd x_1 \int_{x_1}^{\infty^+} \dd x_2 F_{c_{s2}}^\nabla(u,v,x_1,x_2) = - \frac{[4v^2-(1+v^2-u^2)^2]^2}{ 32 u^2v^2} \times  \\
    & \mathrm{Re}\left[ 
    \int_{x_0}^{\infty^+}\dd x_1 \frac{e^{2 i x_1} (i + x_1) (i + c_{s2} u x_1) (1 - i c_{s2} v x_1) (1 + 
     c_{s2} (-i (u + v) + c_{s2} u v x_1) (-2 i - x_1 + c_{s2} (u + v) (i + x_1)))}{(-1 + 
     c_{s2} (u + v))^2 x_1^3}\right] \nonumber \\
     & \quad\quad\quad\quad\quad\quad\quad\quad\quad\quad\quad\quad\quad\quad\quad  \underset{x_0 \rightarrow 0}{=} - \frac{[4v^2-(1+v^2-u^2)^2]^2}{ 32 u^2v^2} \times \nonumber \\
    & \left[\frac{1}{2x_0^2} + \frac{-4 - 8 c_{s2}^4 u^2 v^2 + 8 c_{s2} (u + v) + 5 c_{s2}^5 u^2 v^2 (u + v) +     6 c_{s2}^3 (u + v) (u^2 + u v + v^2) - 
    2 c_{s2}^2 (5 u^2 + 4 u v + 5 v^2)}{(4 (-1 + c_{s2} (u + v))^2} \right] \nonumber \,,
\end{align}
where it is important to use the explicit $i\epsilon$-prescription to eliminate the contributions from the infinite past, and to take the real part to cancel a $1/x_0$ divergence.
We then compute the second contribution, which is actually just one integral squared:
\begin{align}
    &\int_{x_0}^{\infty^+} \dd x_1 \int_{x_0}^{\infty^-} \dd x_2 F_{c_{s2}}^\square(u,v,x_1,x_2) = \frac{1}{2} \frac{[4v^2-(1+v^2-u^2)^2]^2}{ 32 u^2v^2} \times  \\
    & \left|
    i e^{i (1 + c_{s2} (u + v)) x_0} \left[
    \frac{c_{s2} (c_{s2} u^2 (1 + c_{s2} v) + v (1 + c_{s2} v) + u (1 + 4 c_{s2} v + c_{s2}^2 v^2))}{(1 + 
     c_{s2} (u + v))^2} + \frac{i}{x_0} - \frac{i c_{s2}^2 u v x_0}{1 + c_{s2} (u + v)}
     \right]\right|^2 \nonumber \\
     & \quad\quad\quad\quad\quad\quad\quad\quad\quad\quad\quad\quad\quad\quad\quad \underset{x_0 \rightarrow 0}{=} \frac{1}{2} \frac{[4v^2-(1+v^2-u^2)^2]^2}{ 32 u^2v^2} \times \nonumber \\
    & \left[\frac{1}{x_0^2} +
    \frac{c_{s2}^2}{(1 + c_{s2} (u + v))^4} \left(-2 u v (1 + c_{s2} (u + v))^3 + (u + v + c_{s2} v^2 + 
      c_{s2} u^2 (1 + c_{s2} v) + c_{s2} u v (4 + c_{s2} v))^2\right)
    \right] \nonumber \,.
\end{align}
It is clear that the sum of these two terms is free of any divergence.

Then, the loop-integral over the internal momentum may be performed explicitly following these steps (intermediate expressions are lengthy and not particularly illuminating, we do not display them here):
\begin{itemize}
    \item Change the variables $(u,v)$ to $(t,s)$, see Eq.~\eqref{eq: t,s variables};
    \item Perform first the integral over the compact domain $s \in [-1,1]$;
    \item Perform the integral over the infinite domain $t \in [0,\infty[$ by using dimensional regularization, introducing a renormalisation scale $\mu$, and focusing on the log-running that we find to be of the form $\mathrm{log}(\mu/k)$;
    
    \item Add the contribution $\propto \mathrm{log}(k/H)$ from the dimensional regularization of the mode functions~\cite{Senatore:2009cf}, thus arriving at a  
    $\mathrm{log}(\mu/H)$.
\end{itemize}
After this last step, we are left with Eq.~\eqref{eq: Ics2 final} which is the final result.

\subsection{One-loop scalar-tensor-tensor bispectrum}
\label{app: loop mixed bispectrum}

We use the in-in formula for the insertion of three vertices
\begin{align}
    \Braket{\hat{\gamma}^\lambda_{\vec{k}_1}\hat{\gamma}^{\lambda^\prime}_{\vec{k}_2}\hat{\zetan}_{\vec{k}_3}}_{1\mathrm{-loop}}^{\Rpi^3} = &\,
    2 \int^0_{-\infty^-} \dd \tau_1 a(\tau_1) \int^{\tau_1}_{-\infty^-} \dd \tau_2 a(\tau_2) \\
    & \times  \left\{
    \mathrm{Im} \left[
    \int^{\tau_2}_{-\infty^-} \dd \tau_3 a(\tau_3)
    \Braket{0|
    \hat{H}_\mathrm{int}(\tau_3)
    \hat{H}_\mathrm{int}(\tau_2)
    \hat{H}_\mathrm{int}(\tau_1)
    \hat{\gamma}^\lambda_{\vec{k}_1}\hat{\gamma}^{\lambda^\prime}_{\vec{k}_2}\hat{\zetan}_{\vec{k}_3}
    |0}
    \right] \right. \nonumber  \\
    & \left. \quad - \mathrm{Im} \left[
    \int^{0}_{-\infty^+} \dd \tau_3 a(\tau_3) 
    \Braket{0|
    \hat{H}_\mathrm{int}(\tau_2)
    \hat{H}_\mathrm{int}(\tau_1)
    \hat{\gamma}^\lambda_{\vec{k}_1}\hat{\gamma}^{\lambda^\prime}_{\vec{k}_2}\hat{\zetan}_{\vec{k}_3}
    \hat{H}_\mathrm{int}(\tau_3)
    |0}
    \right] \right\} \nonumber \,,
\end{align}
where $\hat{H}_\mathrm{int}=\hat{H}_\mathrm{int}^{\gamma^2\Rpi}+\hat{H}_\mathrm{int}^{\Rpi^3}$.
For each of the two last lines in the above equation, there exist three unequivalent contributions depending on the choice of which of the $\hat{H}_\mathrm{int}(\tau_i)$ is the scalar cubic interaction.
For example if this $\tau_i$ is chosen to be $\tau_3$ in the first of these two lines, one recovers Eq.~\eqref{eq: first term of one-loop bispectrum}.
The five other contributions are straightforward to write from the above equation, so we do not display them all at this stage.

The next steps consist in:
\begin{itemize}
    \item inserting the interaction Hamiltonians;
    \item performing all possible Wick contractions; % while identifying the equivalent ones;
    \item introducing the dimensionless time variables $x_i = - k_1 \tau_i$,
\end{itemize}
after which we arrive at
\begin{align}
    B^{\gamma\gamma\zeta_n}_{1\mathrm{-loop}}(k_1,k_2,k_3)
    &=\frac{\sum_{\lambda,\lambda^\prime}\Braket{\hat{\gamma}^\lambda_{\vec{k}_1}\hat{\gamma}^{\lambda^\prime}_{\vec{k}_2}\hat{\zetan}_{\vec{k}_3}}_{1\mathrm{-loop}}^{\Rpi^3}}
    {(2\pi)^3\delta^{(3)}\left(\vec{k}_1+\vec{k}_2+\vec{k}_3\right) } \nonumber \\
    &=- 24 \pi^2 O_1 \epsilon \frac{\Mp^2}{H^2} \mathcal{P}_{\Rpi}
    \times P_{\Rpi\zetan} (k_3) \times \frac{P_\gamma^{\mathrm{nl}}(k_2)}{I_{c_{s2}}} \times J_{c_{s2}} \,.
\end{align}
This equation is formally the same as  Eq.~\eqref{eq: one-loop BS}, but $J_{c_{s2}}$ is not yet in its simplest form.
In particular, note that the squeezed limit has not been taken yet, so that $J_{c_{s2}}$ reads:
\begin{align}
    J_{c_{s2}} &= \frac{1}{2\pi} \sum_{\lambda,\lambda^\prime}
    \int  \dd^3 \left(\vec{q}/k_1\right)
    \frac{Q_\lambda(\vec{k}_1,\vec{q})}{k_1^2}
    \frac{Q_{\lambda^\prime}(\vec{k}_2,\vec{k}_3+\vec{q})}{k^2}
    \frac{k_1^9}{q^3|\vec{q}-\vec{k}_1|^3|\vec{q}+\vec{k}_3|^3}  \\
    & \quad \times
    \sum_{i=1}^3
    \int_0^{\infty^+} \dd x_1
    \int_{x_1}^{\infty^+} \dd x_2
    \left[
    \int_{x_2}^{\infty^+} \dd x_3 g^{\nabla,i}_{c_{s2}}(x_i) +
    \int_0^{\infty^-} \dd x_3 g^{\square,i}_{c_{s2}}(x_i) \right] \,. \nonumber
\end{align}
We report here the explicit form of the first of the six  $g^{\nabla/\square,i}$ terms:

\begin{align}
    g^{\nabla,i}_{c_{s2}} =& \,
    \left(x_1 x_2 x_3^2\right)^{-2} \mathrm{Im}\left\{
    (1-ix_1)(1+ic_{s2} ux_1)(1+ic_{s2} v x_1) \times \left(1-i\frac{k_2}{k_1} x_2\right)(1-ic_{s2} u x_2)(1 + ic_{s2} \tilde{v} x_2) \right. \nonumber \\ 
    & \times 
    \left[c_{s2} v^2 x_3^2 - 3 (1-i v c_{s2} x_3)  \right]
    \left[c_{s2} \tilde{v}^2 x_3^2 - 3 (1-i \tilde{v} c_{s2} x_3)  \right]
    \left[c_{s2} \frac{k_3^2}{k_1^2} x_3^2 - 3 \left(1-i \frac{k_3}{k_1} c_{s2} x_3\right)  \right] \nonumber \\
    & \times 
    \left.\mathrm{exp}\left[
    -i x_1(c_{s2}u+c_{s2}v-1)
    +i x_2\left(\frac{k_2}{k_1}+c_{s2}u-c_{s2}\tilde{v}\right)
    +i c_{s2} x_3 \left( v + \tilde{v} + \frac{k_3}{k_1}\right)
    \right]\right\}   \,,
\end{align}
where we have introduced the following variables for the internal momenta in the loop:
\begin{align}
    v&=\frac{q}{k_1} \,, \quad 
    u=\frac{|\vec{q}-\vec{k}_1|}{k_1} \,, \quad  \text{and} \quad
    \tilde{v}=\frac{\left|\vec{q}+\vec{k}_3\right|}{k_1} \,.
\end{align}
In the squeezed limit, $k_3 = k_L \ll k_S = k_1 \simeq k_2$ several simplifications occur: $k_2 /k_1 \simeq 1\,,\, k_3/k_1 \ll 1\,,\, \tilde{v} \simeq v$ etc.
After using the same change of variables as in the one-loop power spectrum above, and after integrating over the internal angle $\phi=\mathrm{cos}^{-1}\left[(\vec{q}\cdot \vec{k}_1)/(q k_1)\right] $, we arrive at Eqs.~\eqref{eq: def Jcs2}--\eqref{eq: first term of Gcs2}.
The other five terms are obtained in the same way, but ought to be computed separately.

\paragraph{Computation of $J_{c_{s2}}$.} Each of the six contributions contain IR divergences in the late-time limit. It is convenient to group together the different contributions already
for intermediate steps of the computation.
For example, keeping the focus on the first contribution:
\begin{align}
    &\int_{x_0}^{\infty^+} \dd x_1 \int_{x_1}^{\infty^+} \dd x_2
    \int_{x_2}^{\infty^+} \dd x_3
    G^{\nabla,1}_{c_{s2}}(u,v,x_i) =  \frac{[4v^2-(1+v^2-u^2)^2]^2}{ 32 u^2v^5}\int_{x_0}^{\infty^+} \dd x_1  \\
    &  \times  \mathrm{Im}
    \left\{
    \sum_{n=-6}^2 a_n^{(1)}(c_{s2},u,v) x_1^{-n} e^{2ix_1}
    +
    \sum_{n=-2}^1 b_n(c_{s2},u,v) x_1^{-n}
    \mathrm{Ei}[i(1 + c_{s2} (u + v)) x_1]
    e^{i[1-c_{s2} (u + v) ]x_1}
    \right\} \nonumber \,,
\end{align}
where $ a_n^{(1)}(c_{s2},u,v),  b_n(c_{s2},u,v)$ are complex numbers.
Although we do not integrate analytically the term containing the function $\mathrm{Ei}(z)=- \int_{-z}^\infty \dd t (e^{-t}/t)$, this term is exactly compensated by the second contribution:
\begin{align}
    &\int_{x_0}^{\infty^+} \dd x_1 \int_{x_1}^{\infty^+} \dd x_2
    \int_{x_2}^{\infty^+} \dd x_3
    G^{\nabla,2}_{c_{s2}}(u,v,x_i) =  \frac{[4v^2-(1+v^2-u^2)^2]^2}{ 32 u^2v^5} \int_{x_0}^{\infty^+} \dd x_1 \\
    & \times \mathrm{Im}
    \left\{
    \sum_{n=-6}^2 a_n^{(2)}(c_{s2},u,v) x_1^{-n}e^{2ix_1}
    -
    \sum_{n=-2}^1 b_n(c_{s2},u,v) x_1^{-n}
    \mathrm{Ei}[i(1 + c_{s2} (u + v)) x_1]
    e^{i[1-c_{s2} (u + v) ]x_1}
    \right\} \nonumber \,,
\end{align}
$a_n^{(2)}(c_{s2},u,v)$ being other complex numbers, such that the sum of the two can be computed.
Adding the third contribution, we find in the small-$x_0$ limit:
\begin{align}
    \int_{x_0}^{\infty^+} \dd x_1 \int_{x_1}^{\infty^+} \dd x_2
    \int_{x_2}^{\infty^+} \dd x_3
    G^{\nabla}_{c_{s2}}(u,v,x_i) \underset{x_0 \rightarrow 0}{=} &  \frac{[4v^2-(1+v^2-u^2)^2]^2}{ 32 u^2v^5} \\
    & \times
    \mathrm{Im}
    \left\{
    \sum_{n=-5}^{0} i^{n+1} g_n^\nabla(c_{s2},u,v) x_0^{-n}
    \right\} \,, \nonumber
\end{align}
$g_n^\nabla(c_{s2},u,v)$ being real numbers, from which it is clear that all divergences of odd order in $x_0$ vanish.
The divergences of even order are then cancelled by taking into account the other three contributions, that read:
\begin{align}
    \int_{x_0}^{\infty^+} \dd x_1 \int_{x_1}^{\infty^+} \dd x_2
    \int_{x_0}^{\infty^+} \dd x_3
    G^{\square}_{c_{s2}}(u,v,x_i) \underset{x_0 \rightarrow 0}{=} &  \frac{[4v^2-(1+v^2-u^2)^2]^2}{ 32 u^2v^5} \\
    & \times
    \mathrm{Im}
    \left\{
    \sum_{n=-5}^{0} - i^{n+1} g_n^\square(c_{s2},u,v) x_0^{-n}
    \right\} \,, \nonumber
\end{align}
where the $g_n^\square(c_{s2},u,v)$ are real numbers.
Crucially, the even, non-zero, order terms are equal in both cases, $\left(g_{-4}^\square,g_{-2}^\square\right) =\left(g_{-4}^\nabla, g_{-2}^\nabla\right)$, such that the total result after performing the time integrals is finite and reads:
\begin{equation}
    J_{c_{s2}}=\,\int_0^\infty \dd v  \int_{|1-v|}^{1+v} \dd u\frac{[4v^2-(1+v^2-u^2)^2]^2}{ 32 u^2v^5}\left(g_0^\nabla(c_{s2},u,v)-g_0^\square(c_{s2},u,v)\right) \,.
\end{equation}

We then proceed as in the tensor power spectrum case and obtain Eq.~\eqref{eq: Jcs2 final} as a final result.

\bibliographystyle{JHEP}
\bibliography{biblio}
\end{document}